\documentclass[12pt,preprint]{aastex}
\shorttitle{Photometric redshifts in the nearby universe}
\shortauthors{D'Abrusco et al.}

\begin{document}

\title{Mining the SDSS archive. I. \\
Photometric redshifts in the nearby universe.}
\author{D'Abrusco Raffaele\altaffilmark{1,6}, Staiano Antonino\altaffilmark{5},
 Longo Giuseppe\altaffilmark{1,3,2}, Brescia Massimo\altaffilmark{2,3},
 Paolillo Maurizio\altaffilmark{1,3}, De Filippis Elisabetta\altaffilmark{2,1}, Tagliaferri Roberto\altaffilmark{4,3}}\email{}
\altaffiltext{1}{Department of Physical Sciences, University of
Napoli Federico II, via Cinthia 9, 80126 Napoli, ITALY}
 \altaffiltext{2}{INAF-Italian National Institute of Astrophysics, via del Parco
Mellini, Rome, I} \altaffiltext{3}{INFN -
Napoli Unit, Dept. of Physical Sciences, via Cinthia 9, 80126,
Napoli, ITALY} \altaffiltext{4}{Department of Mathematics and
Applications, University of Salerno, Fisciano, ITALY}\altaffiltext{5}{Department of Applied
Science, University of Napoli "Parthenope",
via A. De Gasperi 5, 80133 Napoli, ITALY}\altaffiltext{6}{Institute of Astronomy, University of Cambridge, Madingley Rd, Cambridge CB4 0HA, UK}

\begin{abstract}

In this paper we present a supervised neural network approach to the
determination of photometric redshifts.
The method, even though of general validity, was fine tuned to
match the characteristics of the Sloan Digital Sky Survey (SDSS) and
as base of 'a priori' knowledge, it exploits the rich wealth of
spectroscopic redshifts provided by this unique survey.
In order to train, validate and test the networks, we used two
galaxy samples drawn from the SDSS spectroscopic dataset, namely:
the general galaxy sample (GG) and the luminous red galaxies
subsample (LRG). Due to the uneven distribution of measured
redshifts in the SDSS spectroscopic subsample, the method consists
of a two steps approach. In the first step, objects are classified
in nearby ($z<0.25$) and distant ($0.25<z<0.50$), with an accuracy
estimated in $97.52\%$. In the second step two different networks
are separately trained on objects belonging to the two redshift
ranges.
Using a standard Multi\ Layer Perceptron operated in a Bayesian
framework, the optimal architectures were found to require 1 hidden
layer of 24 (24) and 24 (25) neurons for the GG (LRG) sample.
The presence of systematic deviations was then corrected by interpolating
the resulting redshifts.

\noindent The final results on the GG dataset give a robust $\sigma _z \simeq 0.0208$
over the redshift range $\left[ 0.01,0.48 \right]$ and
$\sigma _z \simeq 0.0197$ and $\sigma _z \simeq 0.0238$
for the nearby and distant samples respectively.
For the LRG subsample we find instead a robust
$\sigma_z \simeq 0.0164$ over the whole range,  and $\sigma _z \simeq 0.0160$,
$\sigma _z \simeq 0.0183$ for the nearby and distant samples respectively.
After training, the networks have been applied to all objects in the SDSS
Table GALAXY matching the same selection criteria adopted to build the base of
knowledge, and photometric redshifts for ca. 30 million galaxies having $z<0.5$
were derived.
A second catalogue containing photometric redshifts for the LRG subsample was
also produced.
Both catalogues can be downloaded at the URL:
http://people.na.infn.it/~astroneural/SDSSredshifts.htm .
\end{abstract}


\maketitle

\section{Introduction}
After the pioneristic work by the Belgian astronomer Vandererkhoven,
who in the late thirties used prism-objective spectra to derive redshift
estimates from the continuum shape and its macroscopic features (notably
the Balmer break at $\sim 4000$ \AA), \citet{baum_1961} was the
first to test experimentally the idea that redshift could be obtained from
multiband aperture photometry by sampling at different wavelengths the
galaxy spectral energy distribution (hereafter SED).
After a period of relative lack of interest, the 'photometric redshifts'
technique was resurrected in the eighties \citep{butchins_1981}, when it
became clear that it could prove useful in two similar but methodologically
very different fields of application:

\noindent i) as a method to evaluate distances when
spectroscopic estimates become impossible due to either poor signal-to-noise
ratio or to instrumental systematics, or to the fact that the
objects under study are beyond the spectroscopic limit \citep[cf.][]{bolzonella_2002};

\noindent ii) as an economical way to obtain, at a relatively low
price in terms of observing and computing time, redshift estimates
for large samples of objects.

\noindent The latter field of application has been widely explored in the last
few years, when the huge data wealth produced by a new
generation of digital surveys, consisting in accurate multiband
photometric data for tens and even hundreds of millions of
extragalactic objects, has become available.
Photometric redshifts are of much lower accuracy then spectroscopic
ones but even so, if available in large number and for statistically
well controlled samples of objects, they still provide a powerful tool
to derive a 3-D map of the universe.
A map which is crucial for a variety of applications among which we
shall quote just a few: to study large scale structure \citep{brodwin_2006};
to constrain the cosmological constants and models (\citealp{blake_2005} and references therein,
\citealp{budavari_2003} and
\citealp{tegmark_2006});
to map matter distribution using weak lensing (\citealp{edmonson} and references therein).

\noindent In this paper we present a new application of neural networks to the
problem of photometric redshift determination and use the method to
produce two catalogues of photometric redshifts: one for $\sim 30$
million objects extracted form the SDSS-DR5 main GALAXY dataset and
a second one for a Luminous Red Galaxies sample.

\noindent The paper is structured as it follows.
In the Sections~\ref{photz} and \ref{method}, we shortly summarize
the various methods for the determination of photometric redshifts,
and the theory behind the adopted model of neural network.
In \S~\ref{thedata}, we describe both the photometric data set
extracted from the SDSS and the base of knowledge used for the
training and test  and, in \S~\ref{experiments} we
discuss the method and present the results of the experiments.
It needs to be stressed that even though finely tailored
to the characteristics of the SDSS data, the method is general and can
be easily applied to any other set provided that a large enough base of
knowledge is available.

\noindent As stressed by several authors, photometric redshift
samples are useful if the structure of the errors is well understood; in
\S~\ref{errors} we therefore present a discussion of both
systematic and random errors and propose a possible strategy to correct
for systematic errors (\S~\ref{interpolation}).
In \S~\ref{catalogues} we shortly describe the two catalogues.
Finally, in \S~\ref{conclusion}, we discuss the results and present
our conclusions.

\noindent This paper is the first in a series of three. In the second one
\citep{brescia_2006} we shall present the catalogue of structures
extracted in the nearby sample using an unsupervised clustering algorithm
working on the three dimensional data set produced from the SDSS data.
In paper III \citep{dabrusco_2006} we shall complement the information contained
in the above quoted catalogues by discussing the statistical clustering of objects in the photometric parameter space.

\section{Photometric redshifts}
\label{photz}
Without entering into too much detail, photometric redshifts methods can
be broadly grouped in a few families: template fitting, hybrid and
empirical methods.

\noindent Template fitting methods are based on fitting a library of template Spectral
Energy Distributions (SEDs) to the observed data, and differ mainly in how
these SEDs are derived and in how they are fitted to the data.
SEDs may either be derived from population synthesis models \citep{bruzual_1993}
or from the spectra of real objects \citep{coleman} carefully selected in order to
ensure a sufficient coverage of the parameter space (mainly in terms of morphological
types and/or luminosity classes).
Both approaches (synthetic and empirical) have had their pro's and con's
widely discussed in the literature, (cf. \cite{koo_1999}, but see also
\cite{fernandez_2001}, \cite{massarotti_2001a},\cite{massarotti_2001b} and
\cite{csabai_2003}).
Synthetic spectra, for instance, sample an 'a priori' defined grid of
mixtures of stellar populations and may either include unrealistic
combinations of parameters, or exclude some unknown cases.
On the other end, empirical templates are necessarily derived from nearby
and bright galaxies and may therefore be not representative of the spectral
properties of galaxies falling in other redshift or luminosity ranges.
Ongoing attempts to derive a very large and fairly exhaustive set of empirical
templates using the SDSS spectroscopic dataset are in progress and will surely
prove useful in a nearby future.

\noindent Hybrid SED fitting methods making use of a combination of both observed and
theoretically predicted SEDs have been proposed with mixed results by several authors
\citep{bolzonella_2000,padmahaban_2005}.

\noindent The last family of methods, id est the empirical ones, can be applied only to
`mixed surveys', id est to datasets where accurate and multiband
photometric data for a large number of objects are supplemented by spectroscopic
redshifts for a smaller but still significant subsample of the same objects.
These spectroscopic data are used to constrain the fit of an interpolating function
mapping the photometric parameter space and differ mainly in the way such interpolation
is performed.
As it has been pointed out by many authors \citep{connolly_1995,csabai_2003},
in these methods the main uncertainty comes from the fact that the fitting
function is just an approximation of the complex relation existing between the
colors and the redshift of a galaxy and by the fact that as soon as the
redshift range and/or the size of the parameter space increase, a single
interpolating function is bound to fail.
Attempts to overcome this problem have been proposed by several authors.
For instance, \cite{brunner_1999}, divided the redshift and color range in
several intervals in order to optimize the interpolation.
\cite{csabai_2003} used instead an improved nearest neighbor method consisting
in finding, for each galaxy in the photometric sample, the galaxy in the training
set which has the smallest distance in the parameter space and then attributing
the same redshift to the two objects.

\noindent More recently, several attempts to interpolate the a priori knowledge provided by
the spectroscopic redshifts have been made using statistical pattern recognition
techniques such as neural networks \citep{tagliaferri_2002,vanzella_2003,firth_2003} and Support
Vector Machines \citep{wadadekar_2004}, with results which will be discussed more in
detail in what follows.

\noindent It has to be stressed that since the base of knowledge is purely empirical ({\it i.e.}
spectroscopically measured redshifts), these methods cannot be effectively applied to
objects fainter than the spectroscopic limit.
To partially overcome this problem, noticeable attempts have been made to build a
'synthetic' base of knowledge using spectral synthesis models, but it is apparent
that, in this case, the uncertainties of the SED fitting and empirical
methods add up.

\noindent In any case, it is by now well established that when a significant
base of knowledge is available, empirical methods outperform template fitting
ones and that the use of the latter should be confined to those case where a suitable
base of knowledge is missing.

\section{The Multi Layer Perceptron}
\label{method}
Neural Networks (hereafter NNs) have long been known to be
excellent tools for interpolating data and for extracting patterns
and trends and since few years they have also digged
their way into the astronomical community
for a variety of applications
(see the reviews \citep{tagliaferri_2003,tagliaferri_2003_2}
and references therein) ranging from star-galaxy separation \citep{donalek_2007},
spectral classification \citep{winter_2004} and photometric redshifts evaluation \citep{tagliaferri_2002,firth_2003}.
In practice a neural network is a tool which takes a set of input
values (input neurons), applies a non-linear (and unknown) transformation
and returns an output.
The optimization of the output is performed by using a set
of examples for which the output value is known a priori.
NNs exist in many different models and architectures but since the relatively low
complexity of astronomical data does not pose special constrains
to any step of the method which will be discussed below
we used a very simple neural model known as {\it Multi-Layer
Perceptron or MLP} which is probably the most widely used
architecture for practical applications of neural networks.

\begin{figure}
   \centering
   \resizebox{\hsize}{!}{\includegraphics{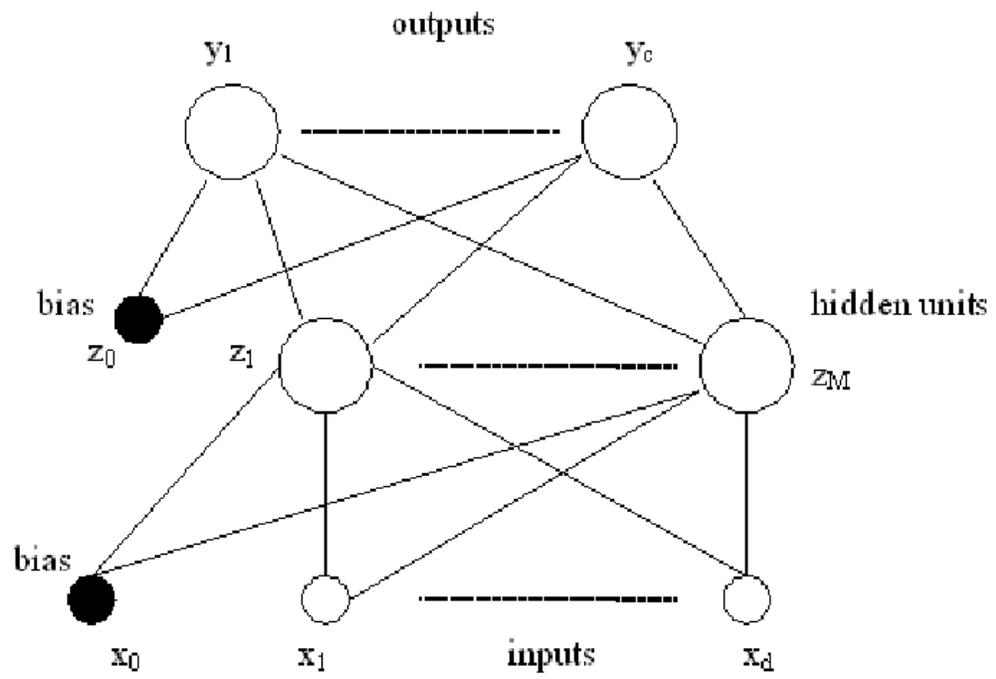}}
     \caption{A schematic representation of the Multi Layer Perceptron.}
\label{Fig:Gam1}
\end{figure}

\noindent In most cases an MLP consists of two layers of adaptive weights with
full connectivity between inputs and intermediate (namely, hidden)
units, and between hidden units and outputs (see Fig.~\ref{Fig:Gam1}).
\clearpage
\clearpage

\noindent Note, however, that an alternative
convention is sometimes also found in literature which counts
layers of units rather than layers of weights, and regards the
input as separate units. According to this convention the network
showed in Fig.~\ref{Fig:Gam1} would be called three-layer network.
However, since the layers of adaptive weights are those which really matter in determining
the properties of the network function, we refer to the former
convention.
\subsection{MLP: the flux of the computation}
The MLP realizes a complex nonlinear mapping from the input to the output
space.
Let us denote the $N$ input values to the network by
$\mathbf{x}=\{x_1,x_2,\ldots,x_d\}$.
The first layer of the network forms a linear combinations of these
inputs to give a set of intermediate activation variables $a_j^{(1)}$
\begin{equation}
a_j^{(1)}=\sum_{i=1}^d w_{ji}^{(1)}x_i+b_j^{(1)}, j=1,\ldots,M
\end{equation}
with one variable $a_j^{(1)}$ associated with each of the $M$
hidden units. Here $w_{ji}^{(1)}$ represents the elements of the
first-layer weight matrix and $b_j^{(1)}$ are the biases
parameters associated with the hidden units. The variables
$a_j^{(1)}$ are then transformed by the nonlinear activation
functions of the hidden layer. Here we restrict attention to
$tanh$ activation functions. The outputs of the hidden units are
then given by
\begin{equation}
z_j=\tanh(a_j^{(1)}), j=1,\ldots,M
\end{equation}
The $z_j$ are then transformed by the second layer of weights and
biases to give the second-layer activation values $a_k^{(2)}$
\begin{equation}
a_k^{(2)}=\sum_{j=1}^M w_{kj}^{(1)}z_j+b_k^{(2)}, k=1,\ldots,c
\end{equation}
where $c$ is the number of output units. Finally, these values are
passed through the output-unit activation function to give output
values $y_k$, where $k=1\ldots,c$. Depending on the nature of the
problem under consideration we have:
\begin{itemize}
    \item for regression problems: a linear activation function, i.e. $y_k=a_k^{(2)}$;
    \item for classification problems: a logistic sigmoidal activation functions applied to
    each of the output independently, i.e.:
    \begin{displaymath}
    y_k=\frac{1}{1+\exp(-a_k^{(2)})},
    \end{displaymath}
\end{itemize}
\subsection{MLP Training Phase}
The basic learning algorithm for MLPs is the so called {\it backpropagation}
and is based on the error-correction learning rule.
In essence, backpropagation consists of two passes through the different
layers of the network: a forward pass and a backward pass.
In the forward pass
an input vector is applied to the input nodes of the network, and
its effect propagates through the network layer by layer. Finally,
a set of outputs is produced as the actual response of the
network. During the backward pass, on the other hand, the weights
are all adjusted in accordance with the error-correction rule.
Specifically, the actual response of the network is subtracted
from a desired (target) response (which we denote as a vector
$\mathbf{t}=\{t_1,t_2,\ldots,t_c\}$) to produce an error signal.
This error signal is then propagated backward through the network.
There are several choices for the form of the error signal to
produce and this choice still depends on the nature of the
problem, in particular:
\begin{itemize}
    \item for regression problems we adopted the sum-of-squares
    error function:
    \begin{displaymath}
    E=\frac{1}{2}\sum_{n=1}^N\sum_{k=1}^c\{y_k(\mathbf{x}^n;\mathbf{w})-t_k^n\}^2;
    \end{displaymath}
    \item for classification problems we used the cross-entropy
    error function:
    \begin{displaymath}
    E=-\sum_{n}\sum_{k=1}^c\{t_k^n\ln y_k^n + (1-t_k^n)\ln(1-y_k^n)\}.
    \end{displaymath}
\end{itemize}

\noindent The weights are adjusted to make the actual response of the
network move closer to the desired response in a statistical
sense. In this work we adopted a computational more efficient
variant of the backpropagation algorithm, namely the quasi-newton
method. Furthermore, we employed a weight-decay regularization
technique in order to limit the effect of the overfitting of the
neural model to the training data, therefore the form of the error
function is:
\begin{displaymath}
\tilde{E}=E+\nu \frac{1}{2}\sum_i w_i^2,
\end{displaymath}
where the sum runs over all the weight and biases. The $\nu$
controls the extents to which the penalty term $\frac{1}{2}\sum_i
w_i^2$ influences the form of the solution.

\noindent It must be stressed that the universal approximation theorem \citep{haykin_1999}
states that the two layers architecture is capable of universal approximation
and a considerable number of papers have appeared in the literature discussing
this property (cf.~\cite{bishop_1995} and reference therein).
An important corollary of this result is that, in the context of a classification
problem, networks with sigmoidal nonlinearities and two layer of weights can
approximate any decision boundary to arbitrary accuracy.
Thus, such networks also provide universal non-linear discriminant functions.
More generally, the capability of such networks to approximate general
smooth functions allows them to model posterior probabilities of
class membership.
Since two layers of weights suffice to implement any arbitrary function,
one would need special problem conditions \citep{duda_2001}
or requirements to recommend the use of more than two layers.
Furthermore, it is found empirically that networks with multiple hidden
layers are more prone to getting caught in undesirable local minima.
Astronomical data do not seem to require such level of complexity
and therefore it is enough to use just a double weights layer,
i.e a single hidden layer.

\noindent As it was just mentioned, it is also possible to train NNs
in a Bayesian framework, which allows to find the more efficient among a
population of NNs differing in the hyperparameters controlling
the learning of the network \citep{bishop_1995}, in the number of
hidden nodes, etc.
The most important hyperparameters being the so
called $\alpha $ and $\beta $.
$\alpha $ is related to the weights of the network and allows to estimate
the relative importance of the different inputs and the selection of the
input parameters which are more relevant to a given task (\textit{Automatic
Relevance Determination}; \cite{bishop_1995}).
In fact,  a larger value for a component of $\alpha $ implies a less meaningful
corresponding weight.
$\beta $ is instead related to the variance of the noise
(a smaller value corresponding to a larger value of the noise) and
therefore to a lower reliability of the network.
The implementation of a Bayesian framework requires several steps:\
initialization of weights and hyperparameters; training the network via a
non linear optimization algorithm in order to minimize the total error
function. Every few cycles of the algorithm, the hyperparameters are
re-estimated and eventually the cycles are reiterated.

\section{The data and the 'base of knowledge'}
\label{thedata}
The Sloan Digital Sky Survey (hereafter SDSS) is an ongoing survey to
image approximately $\pi$ sterad of the sky in five photometric bands
$(u,g,r,i,z)$ and it is also the only survey so far to be complemented
by spectroscopic data for $\sim 10^6$ objects (cf. the SDSS webpages at
http://www.sdss.org/ for further details).
The existence of such spectroscopic subset (hereafter SpS), together
with the accurate characterization of biases and errors
renders the SDSS an unique and ideal playing ground on which
to train and test most photometric redshifts methods.

\noindent Several criteria may be adopted in extracting galaxy data from the SDSS
database \citep{yasuda_2001}.
We preferred, however, to adopt the standard SDSS criterium and
use the GALAXY table membership. The data used in this work were
therefore extracted from the SDSS catalogues. More in particular,
the spectroscopic subsample (hereafter SpS), used for training and
testing purposes, was extracted from the Data Release 4 (hereafter
DR4; cf. \cite{adelman_2005}) . While this work was in progress
the Data Release 5 (DR5) was made publicly available. Thus, the
photometric data used to produce the final catalogues were derived
from the latter data. We wish to stress that this extension of the
dataset was made possible by the fact that the properties of the
DR5 are the same of the DR4 except for a wider sky coverage.

\noindent In this paper we made use of two different bases of knowledge
extracted from the SpS of the DR4:
\begin{itemize}
\item The {\it General Galaxy Sample  or GG}: composed of $445,933$ objects with $z<0.5$
matching the following selection criteria: dereddened magnitude in $r$ band, $r<21$;
$mode=1$ which corresponds to primary objects only in the case of deblended sources.
\item The {\it Luminous Red Galaxies sample or LRG}: composed of $97,475$ red luminous galaxies
candidates having spectroscopic redshift $<0.5$.

\noindent The SDSS spectroscopic survey \citep{eisenstein_2001a} was planned in
order to favour the observation of the so called Red Luminous Galaxies
or LRGs which are expected to represent a more homogeneous population
of luminous elliptical galaxies which can be effectively used to trace
the large scale structures \citep{eisenstein_2001a}.
We therefore extracted from the SDSS-DR4 all objects matching the
above listed criteria and, furthermore, flagged as
$primTarget='TARGET\_GALAXY\_RED'$.
\end{itemize}
LRGs are of high cosmological relevance since they are both
very luminous (and therefore allow to map the universe out to large distances), and
clearly related to the cosmic structures (being preferably found in clusters).
Furthermore, their spectral energy distribution is rather uniform,
with a strong break at $4000$ \AA\ produced by the superposition of
a large number of metal lines \citep{schneider_1983,eisenstein_2003}.
LRGs are therefore an ideal target to test the validity of photometric redshift algorithms
(see for instance:
\cite{hamilton_1985}, \cite{gladders_2000}, \cite{eisenstein_2001a}, \cite{willis_2001}
and \cite{padmahaban_2005}).
The selection of LRG objects was performed using the same criteria extensively described
in \cite{padmahaban_2005} and, given the rather lengthy procedure, we refer to that
paper for a detailed description of the cuts introduced in the parameter space.

\noindent Since it is well known that photometric redshift estimates depend on the morphological type,
age, metallicity, dust, etc. it has to be expected that if some morphological parameters are
taken into account besides than magnitudes or colors alone, estimates of photometric redshifts
should become more accurate.
Such an effect was for instance found  by \cite{tagliaferri_2002,vanzella_2003}.

\noindent In order to be conservative and also because it is not always simple to
understand which parameters might carry relevant information, for each
object we extracted from the SDSS database not only the photometric data
but also the additional parameters listed in Table~\ref{parameters}.

\noindent These parameters are of two types: those which we call 'features'
(marked as $F$ in Table~\ref{parameters}), are parameters which potentially
may carry some useful information capable to improve the accuracy of photometric
redshifts, while those named 'labels' (marked as $L$) can be used to better
understand the biases and the characteristics of the 'base of knowledge'.

\noindent For what magnitudes are concerned, and at a difference with other groups
who used the $modelMag$, we used the so called dereddened magnitudes (\emph{dered}),
corrected for the best available estimate of the SDSS photometric zero-points:
$$\Delta(u,g,r,i,z)=(-0.042, 0.036, 0.015, 0.013, -0.002)$$
as reported in
\cite{padmahaban_2005}. It has to be stressed, however that
such corrections are of little relevance for empirical methods since
they affect equally all data sets.
\clearpage
\begin{deluxetable}{llll}
\tabletypesize{\footnotesize}
%\begin{deluxetable*}{llll}
\tablecolumns{5}
\tablewidth{0pt}
\tablecaption{List of the parameters extracted from the SDSS database and used in the experiments.}
\tablehead{
\colhead{N}  &\colhead{Parameter}  &\colhead{ }&\colhead{F/L}\\
}
\startdata
&  objID & SDSS identification code & -- \\
&  ra    & right ascention (J2000)  & -- \\
&  dec   & declination (J2000)      & -- \\
1 &  petroR50$_i$ & 50 $\%$ of Petr. rad. in the $i$-th band, $i=u,g,r,i,z$& F \\
2 &  petroR90$_i$ & 90 $\%$ of Petr. rad. in the $i$-th band, $i=u,g,r,i,z$& F\\
3 &  dered$_i$ & dered. mag.  in the $i-th$ band, $i=u,g,r,i,z$& F\\
4 &  lnLDeV$_r$ & log likelihood for De Vaucouleurs profile, r band& F\\
5 &  lnLExp$_r$ & log likelihood for exponential profile, r band   & F\\
6 &  lnLStar$_r$ & log likelihood for PSF profile, r band          & F\\
  $z$ & spectroscopic redshift & &L\\
  specClass & spectral classification index& &L \\
\enddata
\tablecomments{Column 1: running number for features only. Column 2: SDSS code. Column 3: short explanation. Column 4: type of parameter, either feature (F) or label (L).}
\label{parameters}
\end{deluxetable}
\clearpage

\noindent Finally we must stress that we impose the condition that the objects had to
be 'primary' ($mode=1$) and detected in all five bands.
The latter condition being required by the fact that all empirical
methods suffer, one way or the other, from the presence of missing data and, to our
knowledge, no clear cut method has been found to overcome this problem.

\subsection{Features selection}
In order to evaluate the significance of the additional features, our first
set of experiments was performed along the same line as described in
\cite{tagliaferri_2002} using a Multi Layer Perceptron with 1 hidden
layer and 24 neurons.
In each experiment, the training, validation and test sets were constructed by
randomly extracting from the overall dataset three subsets, respectively containing
60\%, 20\% and 20\% of the total amount of galaxies.

\noindent On the sample, we run a total of $N+1$ experiments.
The first one was performed using all features,
while the other $N$ were performed taking away the $i-th$ feature with $i=1, ..., N$.
For each experiment, following \cite{csabai_2003}, we used the test set to evaluate the robust variance $\sigma_3$
obtained by excluding all points whose dispersion is larger then 3$\sigma$
(see \S~\ref{catastrophic}).
The values are listed in Table~\ref{feature_sel}.
\clearpage
\begin{deluxetable}{ll}
\tabletypesize{\scriptsize}
\def\arraystretch{1.0}
\tablecolumns{2}
\tablewidth{0pt}
\tablecaption{Results of the feature significance estimation.}
\tablehead{
\colhead{Parameters}  &\colhead{$\sigma_3$}
}
\startdata
all            & 0.0202\\
all but 1      & 0.0209\\
all but 2      & 0.0213\\
all but 4 \& 5 & 0.214\\
all but 6      & 0.215\\
only magnitudes&0.0199\\
\enddata
\tablecomments{Column 1: features used. Features are
numbered as in Table~\ref{parameters}. Column 2: robust sigma of the residuals. }
\label{feature_sel}
\end{deluxetable}
\clearpage
\noindent As it can be seen, the most significant parameters are the magnitudes (or the colors).
Other parameters affect only the third digit of the robust sigma and, due to the large
increase in computing time during the training phase (which scales as $N^2$, where $N$
is the number of input features) and to avoid loss of generality of higher redshifts,
where additional features such as the Petrosian radii are either impossible to measure or
affected by large errors, we preferred to drop all additional features and use only
the magnitudes. The fact that on the contrary of what was found in \cite{vanzella_2003}
and \cite{tagliaferri_2002} additional features do not play a significant role may be understood
as a consequence of the fact that in this work the training set is much larger and more complete
than in these earlier works and therefore the color parameter space is (on average, but see below)
better mapped.
\section{The evaluation of photometric redshifts}
\label{experiments}
One preliminary consideration: as it was first pointed out by \cite{connolly_1995},
when working in the near and intermediate
redshift universe ($z <1$), the most relevant broad band features are
the Balmer break at $4000$ \AA \ and the shape of the continuum in the near UV.
Near IR bands become relevant only at higher redshift and this is the main
reason why we decided to concentrate on the near universe ($z<0.5$), where
the SDSS optical bands provide enough spectral coverage.

\begin{figure}
   \resizebox{\hsize}{!}{\includegraphics[angle=90]{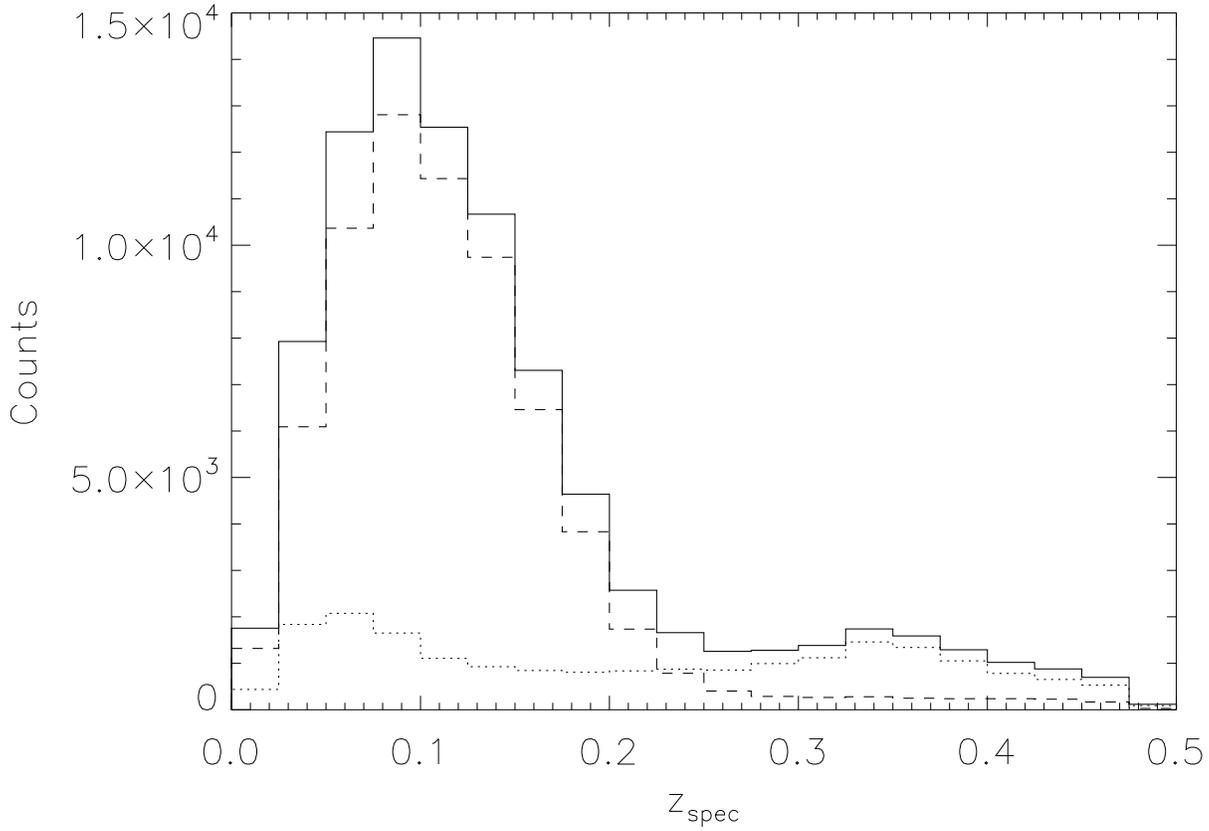}}
\caption{Distribution of redshifts in the SpS sample. Solid line: GG sample.
Dashed line: non-LRG sample.
Dotted line: LRG sample (see text for details).
Notice the sharp drop at $z\sim 0.25$.}
\label{Fig:hist1}
\end{figure}

\noindent One additional reason comes from the redshift distribution of the objects in the
SpS-DR4 shown in Fig.~\ref{Fig:hist1} (solid line).
As it can be clearly seen, the histogram presents a clear discontinuity
at $z\simeq 0.25$ ($86\%$ of the objects have $z <0.25$ and only $14\%$
are at a higher redshift) and in practice no objects are present for $z>0.5$.
\clearpage

\begin{figure}
\resizebox{\hsize}{!}{\includegraphics[angle=90]{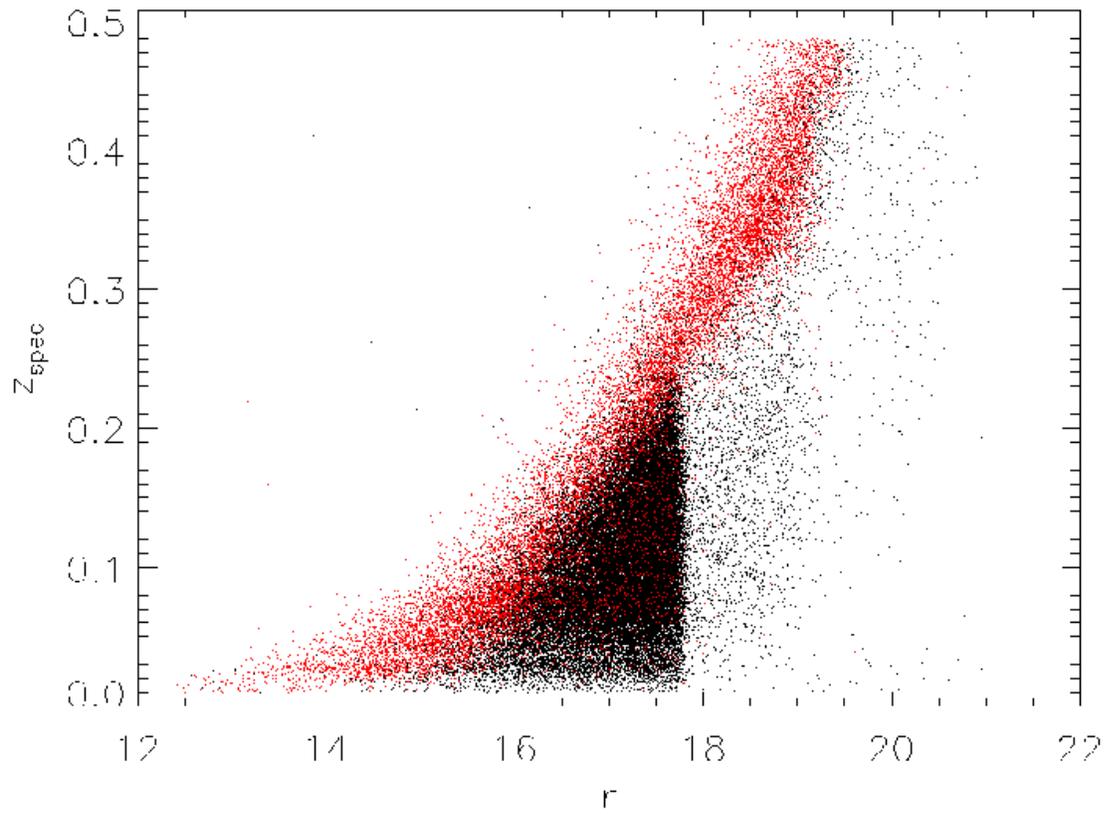}}
\caption{Distribution of the objects in the GG sample versus the $r$ magnitude.
We plot the LRG objects as red dots.} \label{Fig:distribution}
\end{figure}
\clearpage
In Fig.~\ref{Fig:hist1} we also plot as dotted line the redshift distribution of the
galaxies in the SpS data set which match the LRG photometric selection
criteria. As it can be seen, within the tail at $z>0.25$ only a very small
fraction (11.4\%) of the objects does not match the LRG selection criteria.
In Fig.~\ref{Fig:distribution} we plot the redshift of objects belonging to
the GG sample against their luminosity
in $r$ band: red dots represent those galaxies which
have been \emph{a posteriori}
identified as LRG.
As it is clearly seen, the overall distribution at redshift $\leq 0.25$ drops
dramatically at $r \sim 17.7$, due to the selection criteria of the spectroscopic SDSS survey.
At higher redshift, namely $z > 025$, the galaxy distribution is dominated by LRGs with few contaminants and extends
to much fainter luminosities.
Nevertheless LRGs are systematically brighter then  GG galaxies all over the redshift
interval $z < 0.50$.

\noindent Such large dishomogeneity in the density and nature of training data, poses severe
constraints on any empirical method since the different weights of samples
extracted in the different redshift bins would lead either to over fitting
in the densest region, or to the opposite effect in the less populated ones.
Furthermore, the dominance of LRGs at $z>0.25$ implies that in this
redshift range the base of knowledge offers a poor coverage of the parameter
space.

\noindent The first problem can be solved by taking into account the fact that,
as shown in \cite{tagliaferri_2002,firth_2003} NNs work properly even
with scarcely populated training sets, and by building a training set
which uniformly samples the parameter space or, in other words,
which equally weights different clusters of points
(notice that in this paper we use the word cluster in the statistical
sense, id est to denote a statistically significant
aggregation of points in the parameter space).
In the present case the dominance of LRGs at high redshifts renders
the parameter space heavily undersampled.

\noindent In fact, as it will be shown in Paper III \citep{dabrusco_2006},
a more detailed analysis of the parameter space shows that at high redshift,
the objects group into one very large structure containing more than 90$\%$
of the data points, plus several dozens of much smaller clusters .

\subsection{The nearby and intermediate redshifts samples}
In order to tackle the above mentioned problems, we adopted a two steps
approach: first we trained a network to recognize nearby (id est with
$z<0.25$) and distant ($z>0.25$) objects, then we trained two separate
networks to work in the two different redshift regimes.
This approach ensures that the NNs achieve a good generalization
capabilities in the nearby sample and leaves the biases mainly
in the distant one.
To perform the separation between nearby and distant objects,
we extracted from the SDSS-4 SpS training, validation and test
sets weighting, respectively, $60\%$, $20\%$ and $20\%$ of the
total number of objects (449,370 galaxies).
The resulting test set, therefore, consisted of 89,874  randomly
extracted objects.
Extensive testing (each experiment was done performing a separate random
extraction of training, validation and test sets) on the network architecture
lead to a MLP with 18 neurons in 1 hidden layer.
This NN achieved the best performances after 110 epochs and the results are
detailed, in the form of a confusion matrix, in Table~\ref{cm1}.

\noindent As it can be seen, this first NN is capable to separate the two classes
of objects with an efficiency of $97.52\%$, with slightly better performances
in the nearby sample ($98.59\%$) and slightly worse in the distant one ($92.47\%$).

\clearpage
\begin{deluxetable}{lcc}
\tabletypesize{\scriptsize}
\def\arraystretch{1.0}
\tablecolumns{3}
\tablewidth{0pt}
\tablecaption{Confusion matrix for the ''nearby-distant'' test set.}
\tablehead{
\colhead{} &\colhead{SDSS nearby}  &\colhead{SDSS far}
}
\startdata
NN nearby & 76498&1096\\
NN far &1135&11145 \\
\enddata
\label{cm1}
\end{deluxetable}
\clearpage

\begin{figure}
\resizebox{\hsize}{!}{\includegraphics[angle=90]{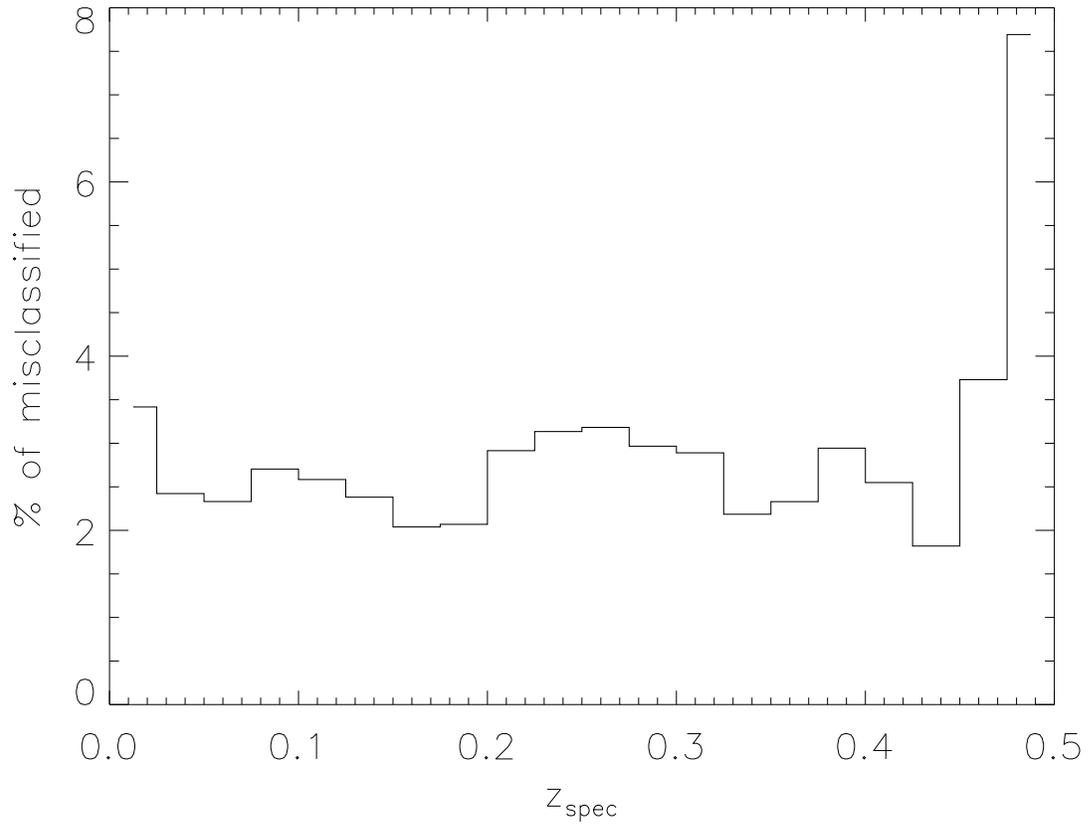}}
\caption{Percentage distribution of misclassified objects of GG
sample normalized to the total number of galaxies in each redshift
bin.} \label{Fig:hist2}
\end{figure}

\noindent In Fig.~\ref{Fig:hist2} we plot against the redshift
the percentage (calculated binning over the redshifts) for the  objects in the test set which were misclassified (id est objects
belonging to the nearby sample which were erroneously attributed to the
distant one and viceversa).
The distribution appears fairly constant from $z_{spec} \sim 0.05$ to $z_{spec} \sim 0.45$,
while higher (but still negligible respect to the total number of objects in the sample) percentages
are found at the extremes.

\clearpage
\clearpage

\noindent Notice that, when using photometric data alone, the absence of training data for $z>0.5$,
does not allow to evaluate the fraction of contaminants having $z>0.5$ which are
erroneously attributed to the distant sample. However, given the adopted cuts in
magnitude, this number may be safely assumed to be negligible.

\subsection{The photometric redshifts}
Once the first network has separated the nearby and distant objects, we can proceed
to the derivation of the photometric redshifts working separately in the two regimes.
Since NNs are excellent at interpolating data but very poor in extrapolating them,
in order to minimize the systematic errors at the extremes of the training redshift
ranges we adopted the following procedure.

\noindent For the nearby sample we trained the network using objects with spectroscopic redshift
in the range $\left[ 0.0,0.27 \right]$ and then considered the results to be reliable
in the range $\left[ 0.01,0.25 \right]$.
In the distant sample, instead, we trained the network over the
range $\left[ 0.23,0.50 \right]$ and then considered the results to be reliable in the
range $\left[ 0.25,0.48 \right]$.

\begin{figure}
   \centering
   {\includegraphics[angle=90,width=10cm]{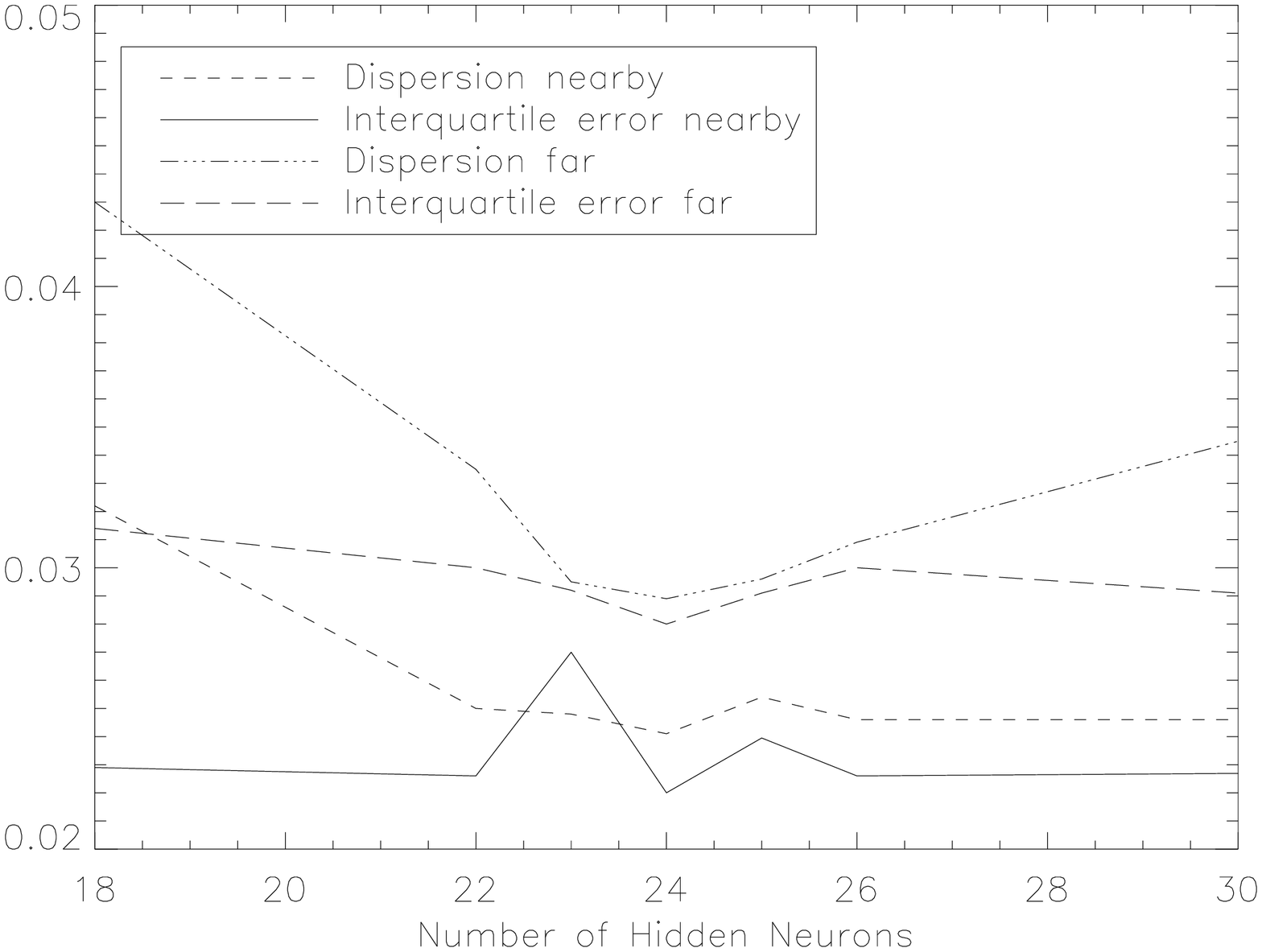}}
   {\includegraphics[angle=90,width=10cm]{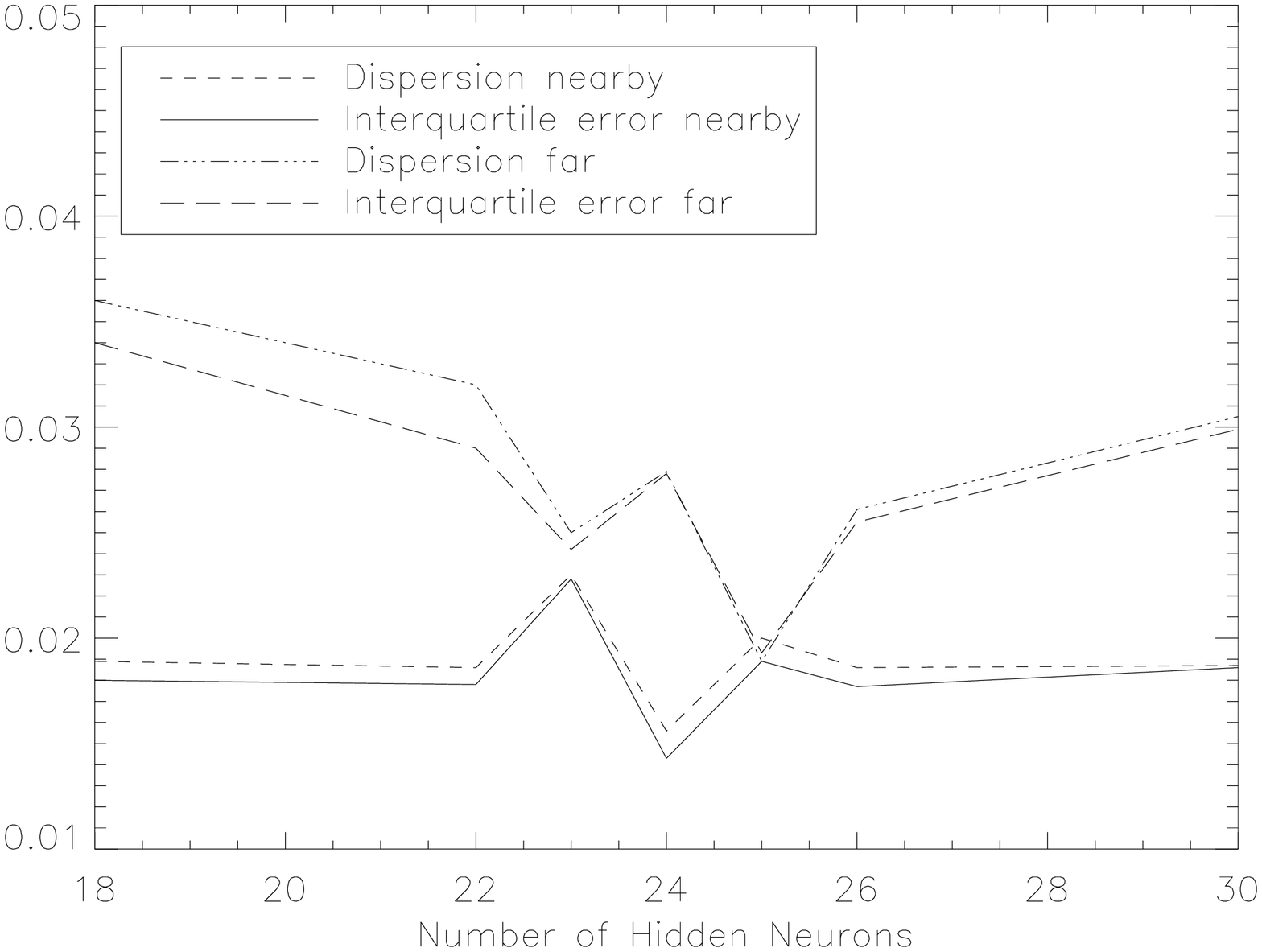}}
     \caption{Upper panel: GG sample, trend of the interquartile
     error and of the robust $\sigma$ as a function of the number N of
     the neurons in the hidden layer.
     The nearby and distant samples are plotted separately.
     Lower panel: the same as above for the LRG sample.}
\label{Fig:neurons}
\end{figure}

\begin{figure}
\centering
   {\includegraphics[angle=90,width=10cm,keepaspectratio]{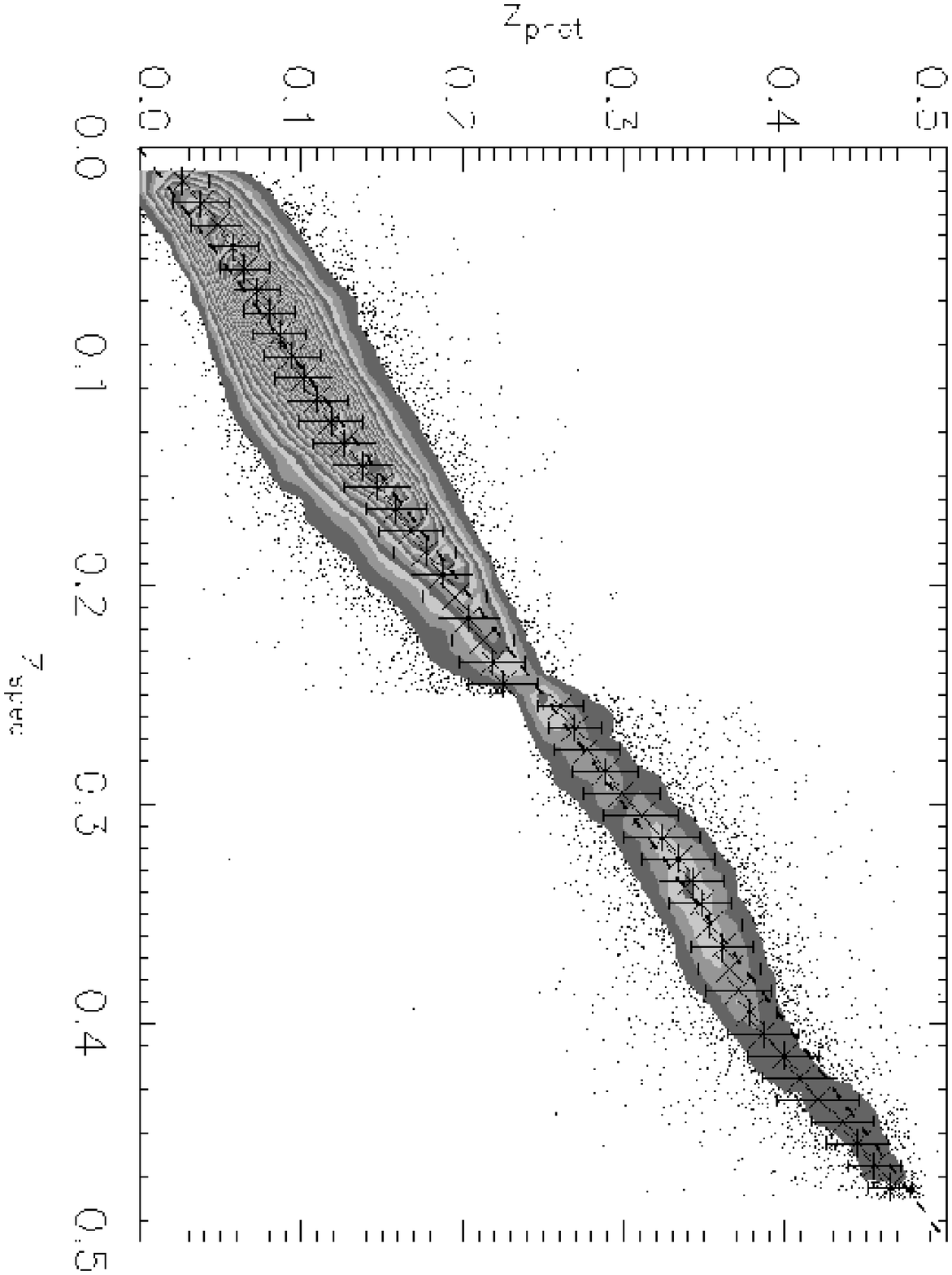}}
   {\includegraphics[angle=90,width=10cm,keepaspectratio]{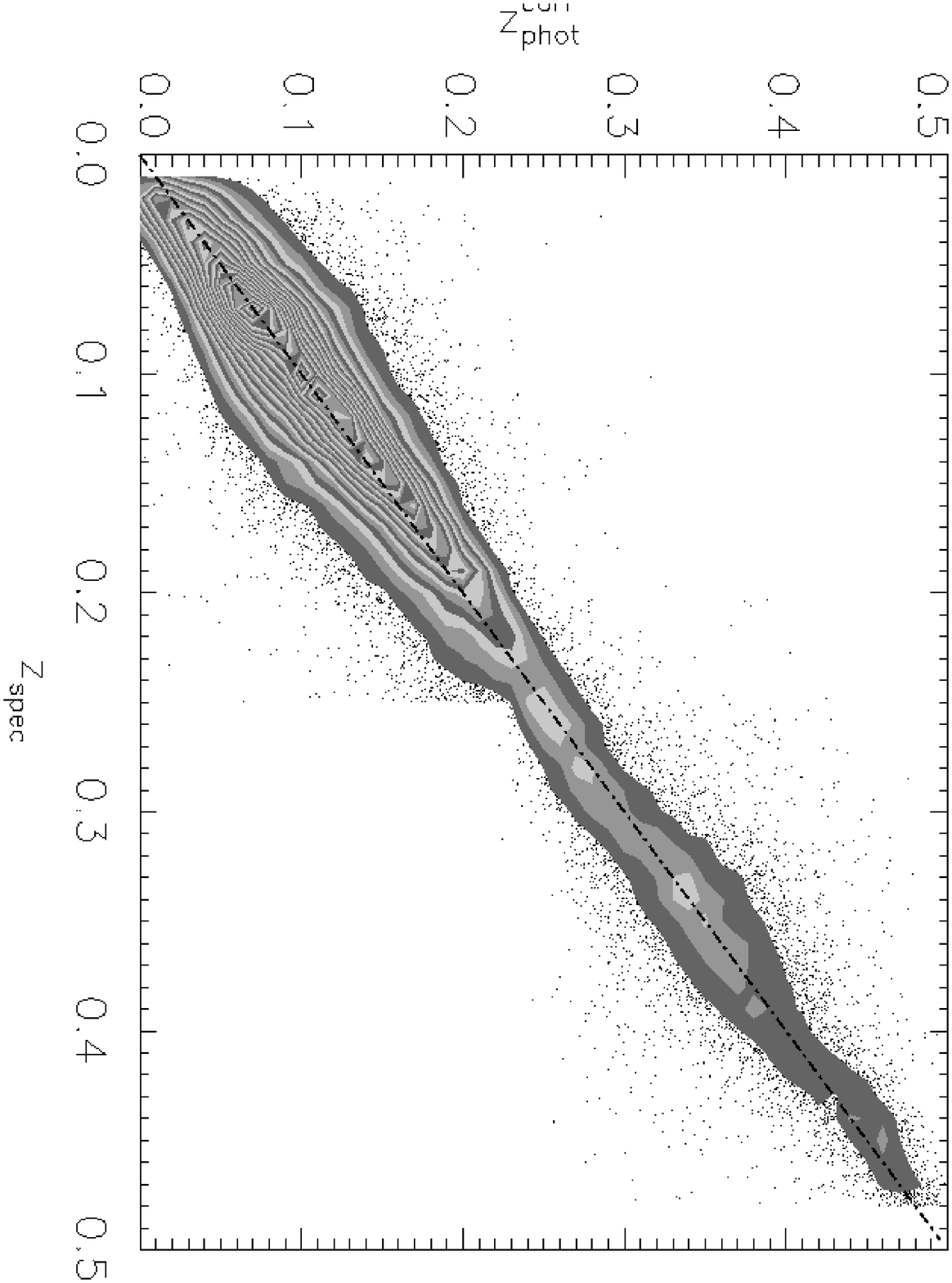}}
     \caption{Upper panel: photometric versus spectroscopic redhifts for
     the objects in the GG test set.
     The continuous lines are iso-density contours increasing with a step of 2 \% of the maximum density.
     The crosses mark the average value of photometric redshifts in a specific spectroscopic
     redshift bin (see text), while the error bars give the robust variance $\sigma_3$.
     Lower panel: same as above after the correction for the systematic trends via interpolation
     (see text).}
\label{Fig:GGS_omc}
\end{figure}

\begin{figure}
\centering
   {\includegraphics[angle=90,width=10cm,keepaspectratio]{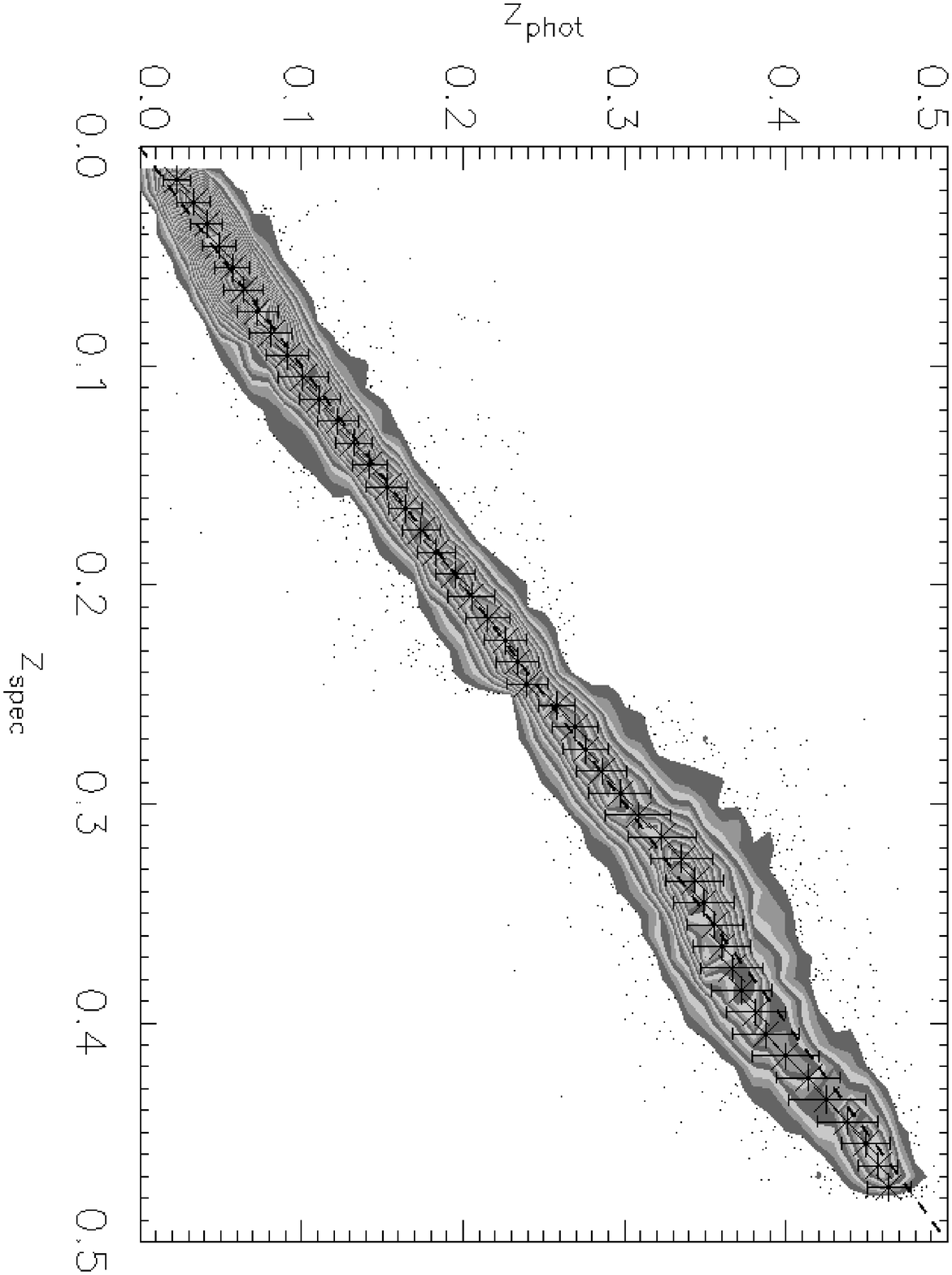}}
   {\includegraphics[angle=90,width=10cm,keepaspectratio]{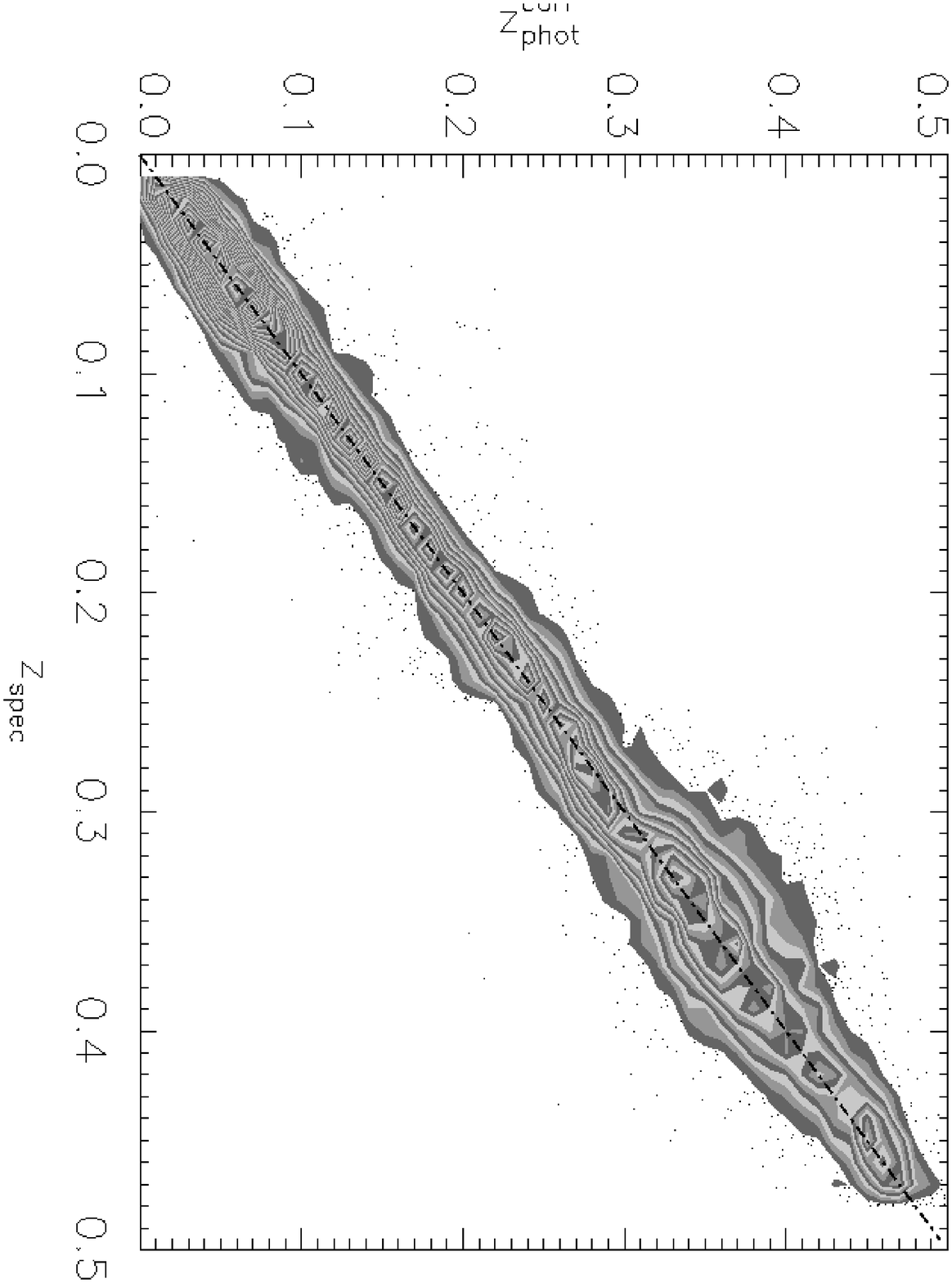}}
\caption{Same as in Fig.~\ref{Fig:GGS_omc} for the LRG sample.}
\label{Fig:LRG_omc}
\end{figure}

\noindent In order to select the optimal NN architecture, extensive testing was made varying the
network parameters and for each test the training, validation and test sets were
randomly extracted from the SpS.
The results of the Bayesian learning of the NNs
were found to depend on the number of neurons in the hidden layer; for the
GG (LRG) sample the performances were best when this parameter was set to 24 for the nearby
sample and for the distant one (24 and 25 respectively for the LRG sample).
In Fig.~\ref{Fig:neurons} we give the trends as a function of the number of hidden neurons, of the interquartile errors and robust dispersion
obtained for the nearby and distant GG samples respectively.
\clearpage
\clearpage
For the GG sample, the best experiment, the robust variance turned out
to be $\sigma_3 = 0.0208$ over the whole redshift range and $0.0197$ and
$0.0245$ for the nearby and distant objects, respectively.
For what the LRG sample is concerned, we obtained $\sigma_3 \simeq 0.0163$ over the whole range,
and $\sigma _3 \simeq 0.0154$ and $\sigma _3 \simeq 0.0189$ for the nearby and distant samples,
respectively.
In the upper panels of Figs.~\ref{Fig:GGS_omc} and~\ref{Fig:LRG_omc} we plot the spectroscopic versus the photometric
redshifts  for the GG and the LRG samples, respectively.
Due to the huge number of points which would make difficult to see the trends in the densest regions,
we preferred to plot the data using isocontours (using a step of 0.02 times the maximum data point density).

\clearpage

\begin{figure}
\centering
   {\includegraphics[width=10cm,angle=90]{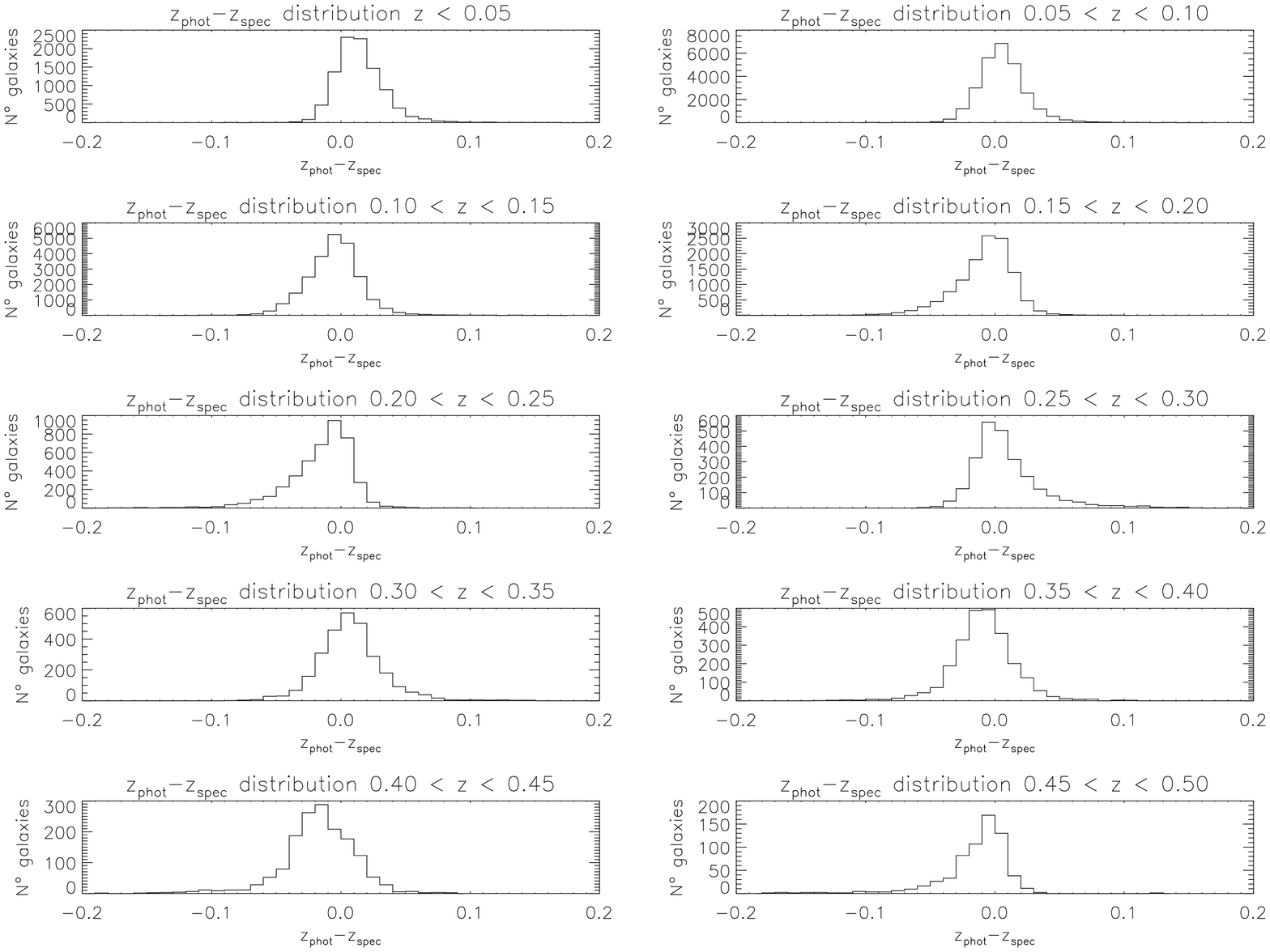}}
\vspace{0.3cm}

   {\includegraphics[width=10cm,angle=90]{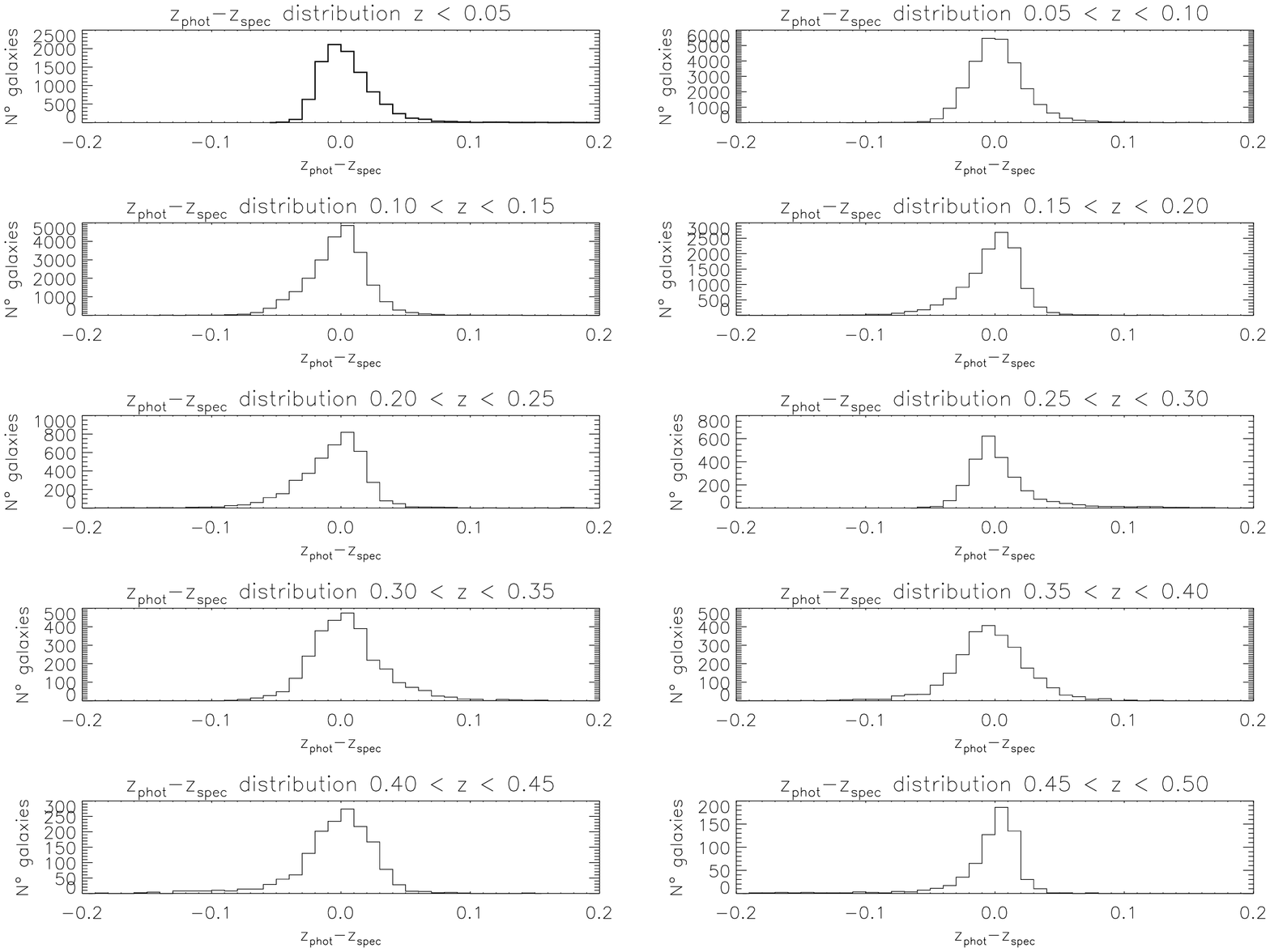}}
   \caption{Histograms of residuals for the GG sample in slices of redshift. Upper panels:
   before the correction. Lower panels: after the correction.}
\label{Fig:errors_slices_GG}
\end{figure}

\begin{figure}
\centering
   {\includegraphics[width=10cm,angle=90]{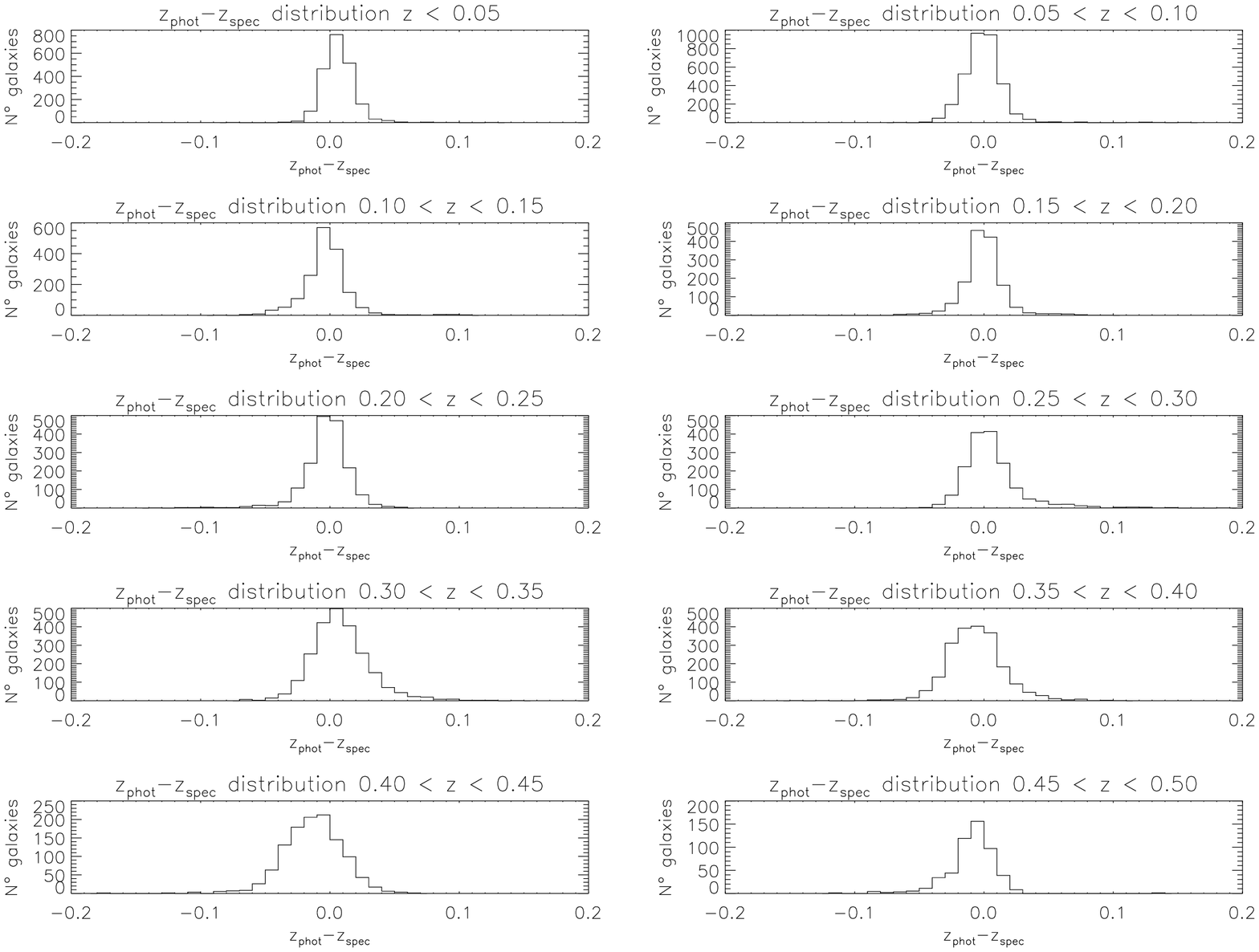}}
\vspace{0.3cm}

   {\includegraphics[width=10cm,angle=90]{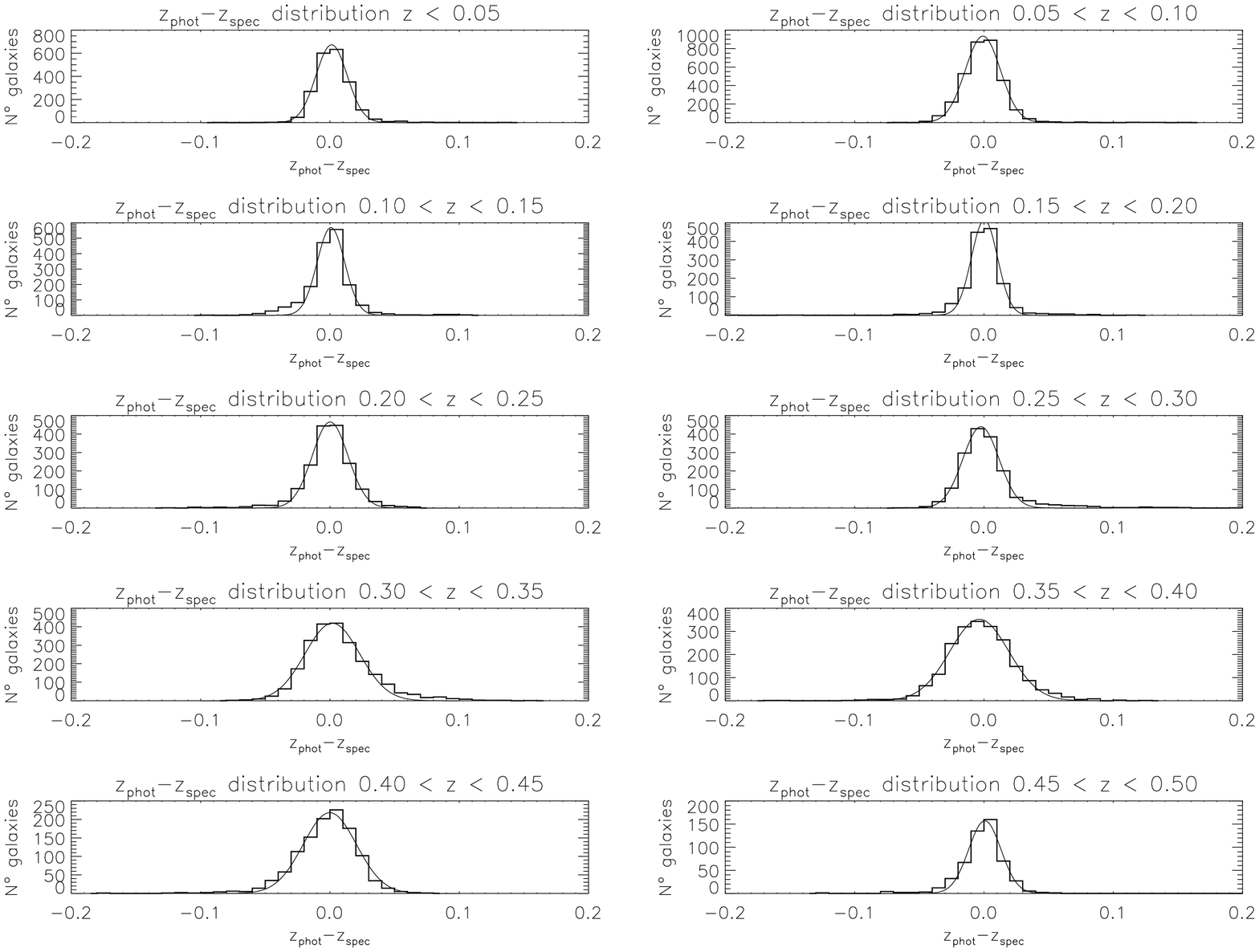}}
   \caption{Same as in previous figure but for the LRG sample.}
\label{Fig:errors_slices_LRG}
\end{figure}

\clearpage
\noindent The mean value of the residuals are $-0.0036$ and $-0.0029$ for the GG
and the LRG samples, respectively.
These figures alone, however, are not very significant since systematic trends are
clearly present in the data as it is shown in Fig.~\ref{Fig:errors_slices_GG}
and in Fig.~\ref{Fig:errors_slices_LRG}, where we plot for each $0.05$ redshift bin the average value
of the photometric redshifts and the robust sigma of the residuals.
\clearpage
\clearpage

\section{Interpolative correction}
\label{interpolation}
The most significant deviations, as it could be expected \citep{connolly_1995},
are clearly visible in the nearby sample for $z<0.1$ and
in the distant sample at $z\sim 0.4$.
The first feature is due to the fact that at low redshifts
faint and nearby galaxies cannot be easily disentangled by
luminous and more distant objects having the same color.
The second one is instead due to a degeneracy in the SDSS
photometric system introduced by a small gap between the $g$ and $r$ bands.
At $z\sim 0.4$, the
Balmer break falls into this gap and its position becomes ill defined
\citep{padmahaban_2005}.

\noindent It needs to be stressed, however, that these trends represent
a rather normal behavior for empirical methods which has already been
explicitly noted in \cite{tagliaferri_2002} and
\cite{vanzella_2003} and is clearly visible (even when it is not explicitly mentioned)
in almost all photometric redshifts data sets \citep{wadadekar_2004} available so far
for the SDSS.

\noindent In order to minimize the effects of such systematic trends, but at the risk of a
slight increase in the variance of the final catalogues we applied to both data sets
an interpolative correction computed separatedly in the two redshift intervals.
We used a ($\chi^{2}$ fitting) to find, separately in each redshift regime,
the polynomials which best fit the average points.
These polynomials (of the fourth and fifth order, respectively)
turned out to be. For the GG sample:
\footnotesize
\begin{eqnarray}
\textsl{P}_{4} &\left[ 0.005,1.570, -12.577, 78.948, -157.961 \right]\\
\textsl{P}_{5} &\left[ 12.15, -178.2, 1039.3, -2959.0, 4135.5, -2271.3 \right]
\end{eqnarray}
\normalsize
and for the LRG sample:
\footnotesize
\begin{eqnarray}
\textsl{P}_{4} &\left[ 0.011,0.885, -1.820, 21.350, -53.159 \right] &\\
\textsl{P}_{5} &\left[ 13.1, -192.5, 1123.3, -3207.2, 4504.5, -2491.6 \right]&
\end{eqnarray}
\normalsize

\noindent Thus, the correction to be applied is:
\begin{equation}
z_{phot}^{corr}=z_{phot}-(z_{phot}^{calc}-z_{spec})
\label{formula_1}
\end{equation}
where $z_{phot}^{calc}=\textsl{P}_{4}(z_{spec})$ for near objects
and $z_{phot}^{calc}=\textsl{P}_{5}(z_{spec})$ for the distant ones.

\noindent Obviously, when applying this method to objects for which
we do not possess any spectroscopic estimate of redshift,
it is impossible to perform the transformation (Eq.~\ref{formula_1})
to correct NNs $z_{phot}$ estimates for systematic trends and we
are obliged to use an approximation.
In other words, we replace the unknown $z_{spec}$ with $z_{phot}$ in the
Eq.~(\ref{formula_1}), obtaining the relation:

\begin{equation}
\tilde{z}_{phot}^{corr}= z_{phot} - (\tilde{z}_{phot}^{calc} -
z_{phot})
\end{equation}

\noindent where $\tilde{z}_{phot}^{calc} = \textsl{P}_{4}(z_{phot})$ or
$\tilde{z}_{phot}^{calc} = \textsl{P}_{5}(z_{phot})$ depending on
the redshift range.

\noindent This is equivalent to assuming that the same NNs $z_{phot}$ distribution
represents, with good approximation, the underlying and unknown $z_{spec}$
distribution.
After this correction we obtain a robust variance $\sigma_{3}=0.0197$
for the GG sample and $0.0164$ for the LRG samples, computed in both cases
over the whole redshift range, and the resulting distributions for the two
samples are shown in the lower panels of Figs.~\ref{Fig:GGS_omc} and ~\ref{Fig:LRG_omc}.
\section{Discussion of the systematics and of the errors}
\label{errors}
As noticed by several authors (see for instance,
\cite{schneider_2006,padmahaban_2005}), while some tolerance can
be accepted on the amplitude of the redshift error, much more
critical are the uncertainties about the probability distribution
of those errors.
This aspect is crucial since \citep{padmahaban_2005}
the observed redshift distribution is related to the true redshift distribution
via a Fredholm equation which is ill defined and
strongly dependent on the accuracy with which the noise can
be modeled.
In this respect, many recent studies on the impact of redshift
uncertainties on various cosmological aspects are available:
dark energy from supernovae studies and cluster number counts
\citep{huterer_2004}; weak lensing
\citep{bernstein_2004,huterer_2006,ishak_2005,ma_2006};
baryon oscillations \citep{zhan_2005,zhan_2006}.
All these studies model the error distribution as Gaussian.

\noindent However, photometric redshift error distributions, due to spectral-type/redshift degeneracies,
often have bimodal distributions, with one smaller peak separated from a larger peak by z of order unity
\citep{benitez_2000,fernandez_2001,fernandez_2002}, or more complex error distributions, as it can be seen
in Fig.~\ref{Fig:errors_slices_GG} within the GG sample.
\clearpage
\begin{figure}
\centering
   \resizebox{\hsize}{!}{\includegraphics[angle=90,width=11cm,keepaspectratio]{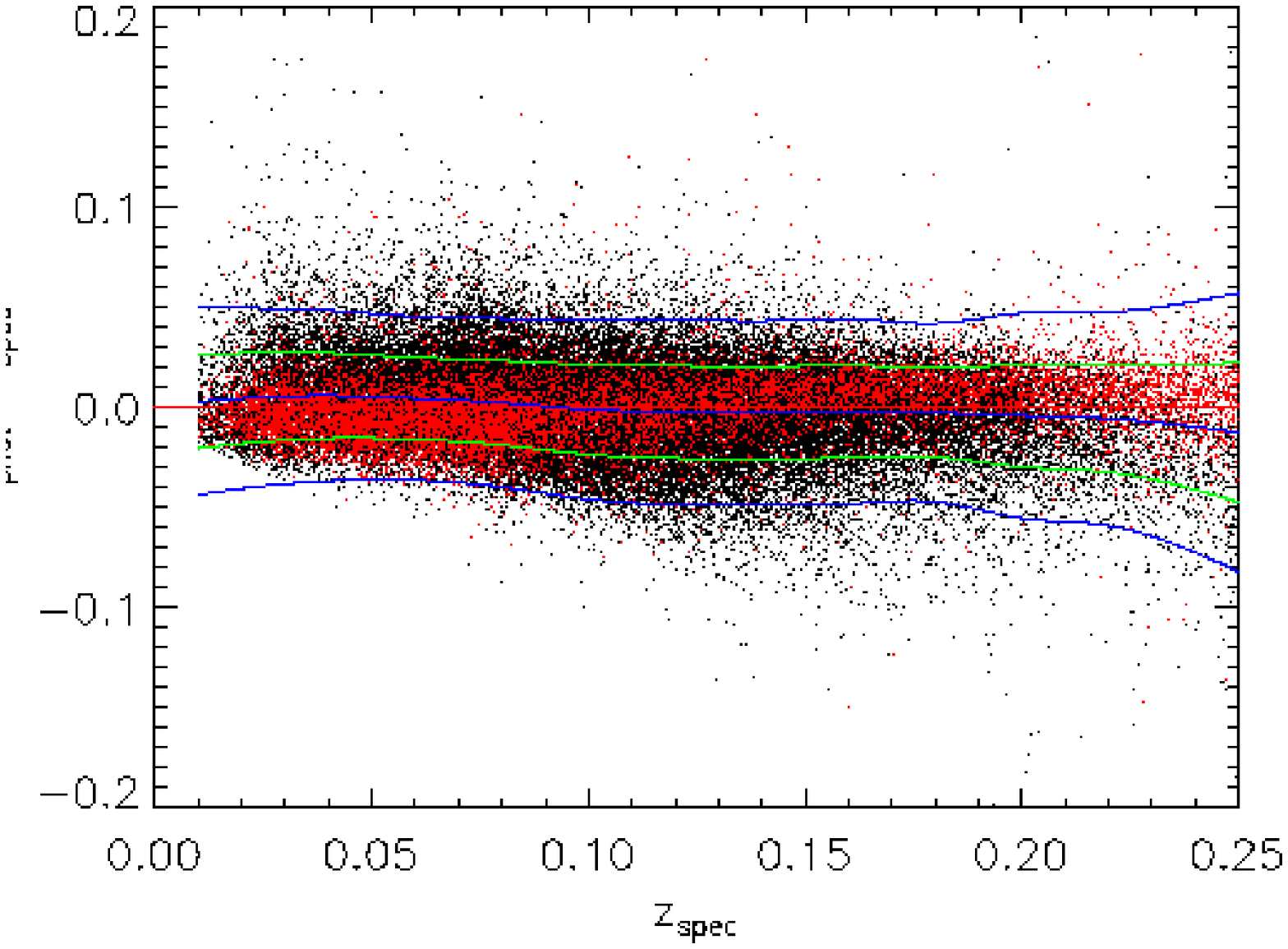}
   \includegraphics[angle=90,width=11cm,keepaspectratio]{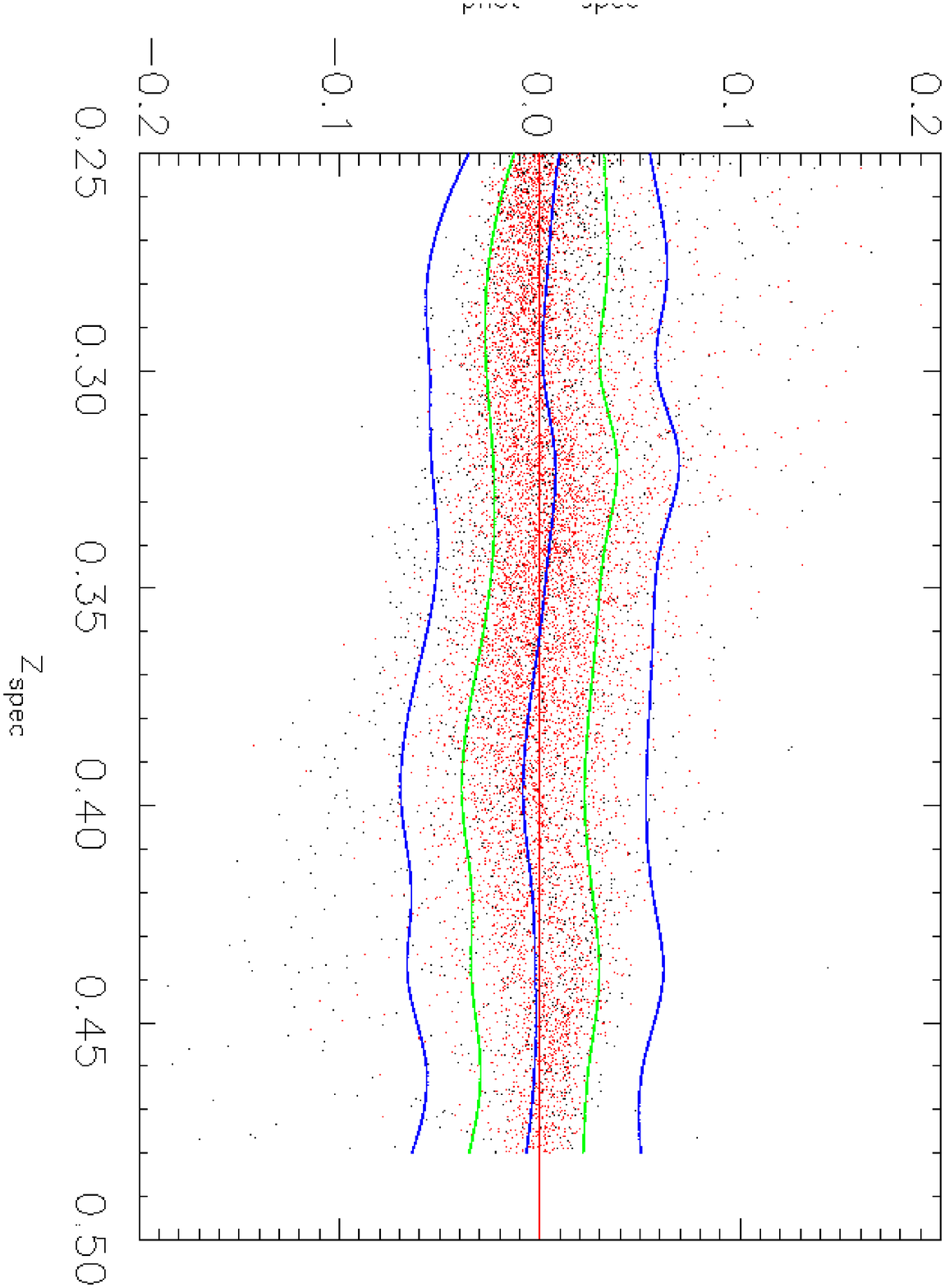}}

\resizebox{\hsize}{!}{\includegraphics[angle=90,width=11cm,keepaspectratio]{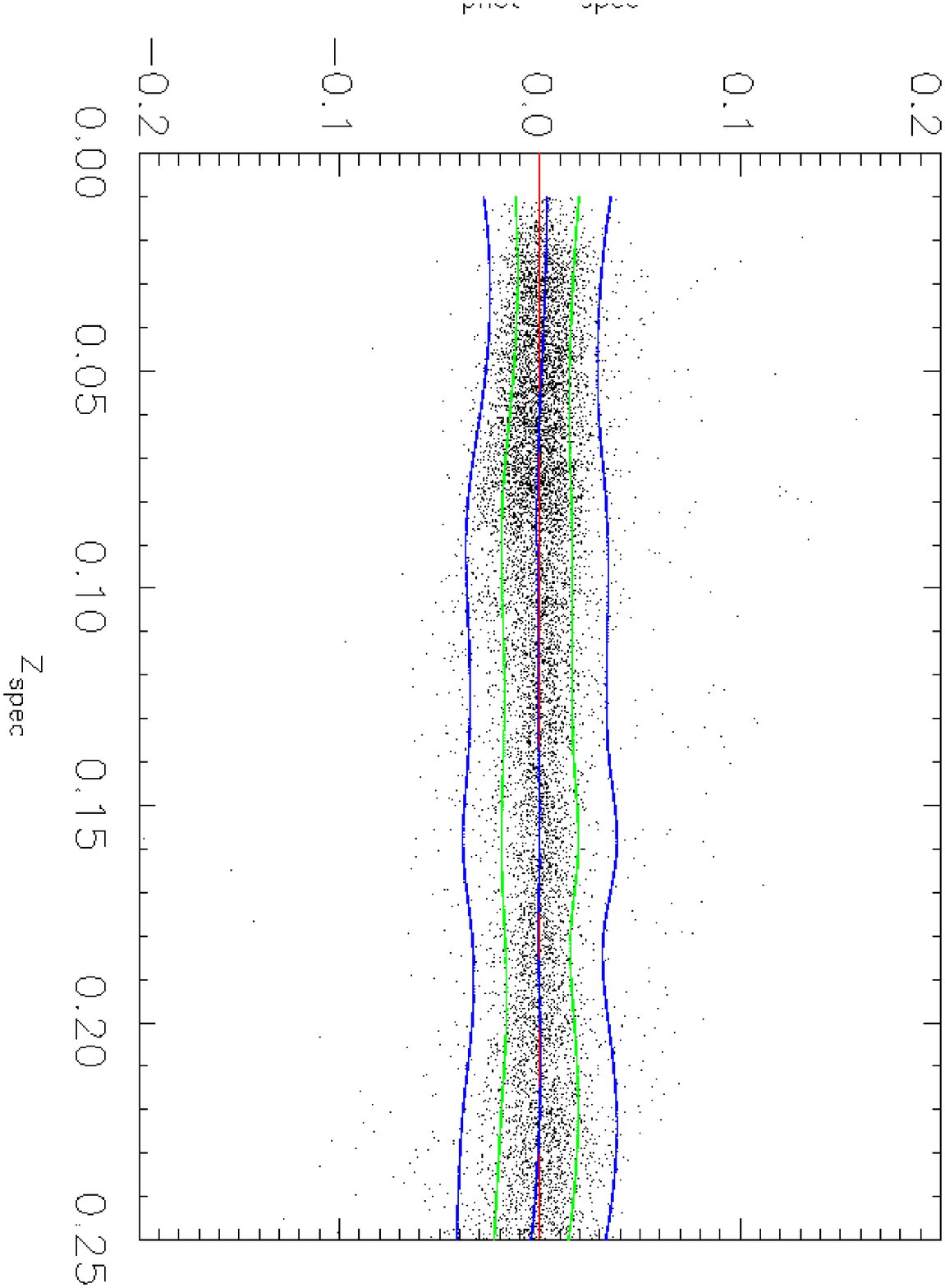}\includegraphics[angle=90,width=11cm,keepaspectratio]{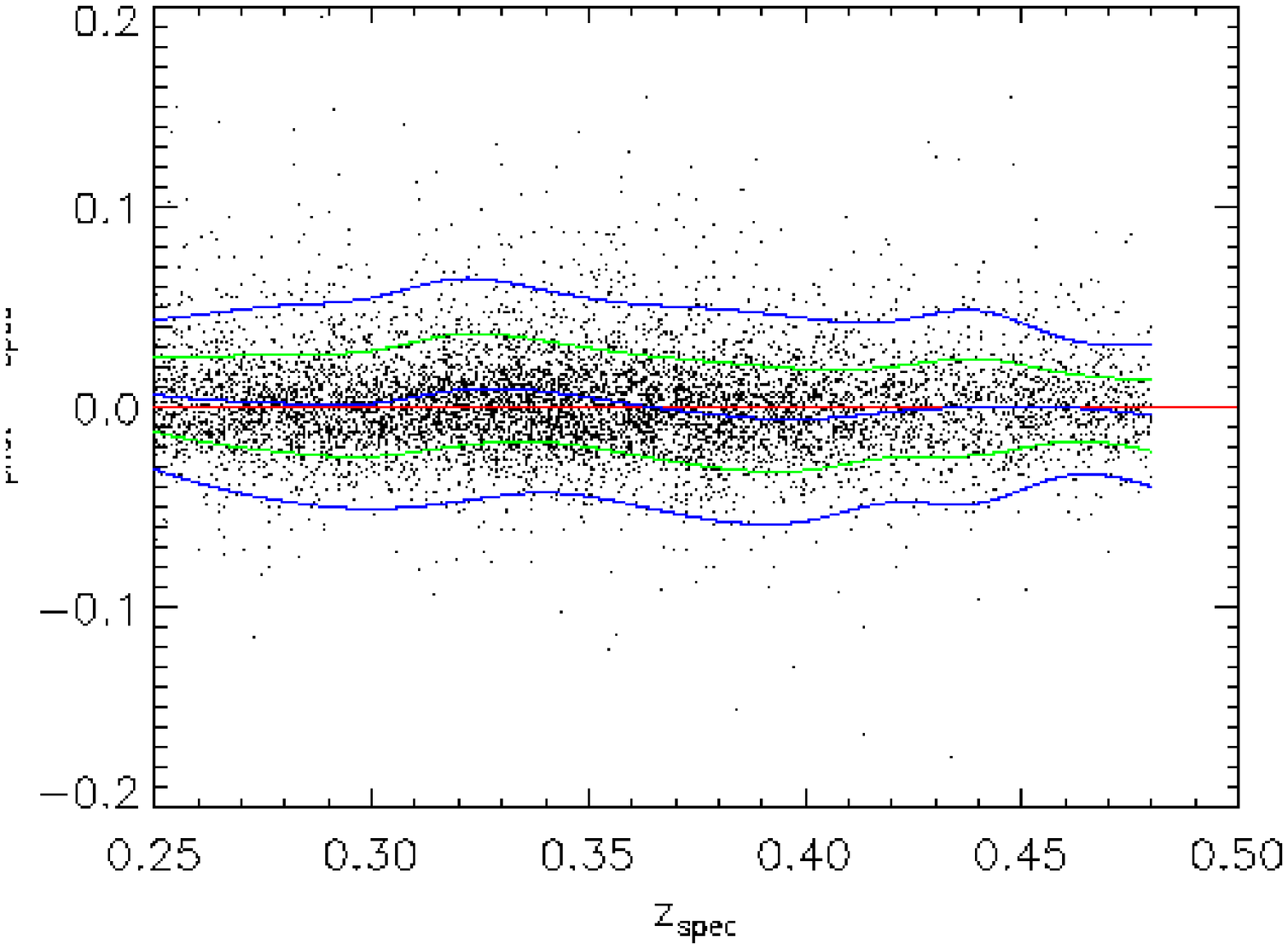}}
     \caption{Distribution of the residuals versus spectroscopic redshift after
     the correction for systematic trends. Upper panels: GG nearby and distant samples.
     Lower panels: LRG nearby and distant samples. The central line marks the
     average value of the residuals. The $1 \ \sigma$ and  $2 \ \sigma$ confidence levels
     are also shown.}
\label{Fig:errors_totale}
\end{figure}
\clearpage

In order to evaluate the robustness of the $\sigma_r$, several instances of
the process were applied to different randomly selected training, validation and test sets
and the robust sigma was found to vary only on the fourth significant digit.
Small differences were found only in the identification of catastrophic objects,
which however did not present any significant variation in their
frequency.

\begin{figure}
\centerline{\includegraphics[angle=90,width=5cm,keepaspectratio]{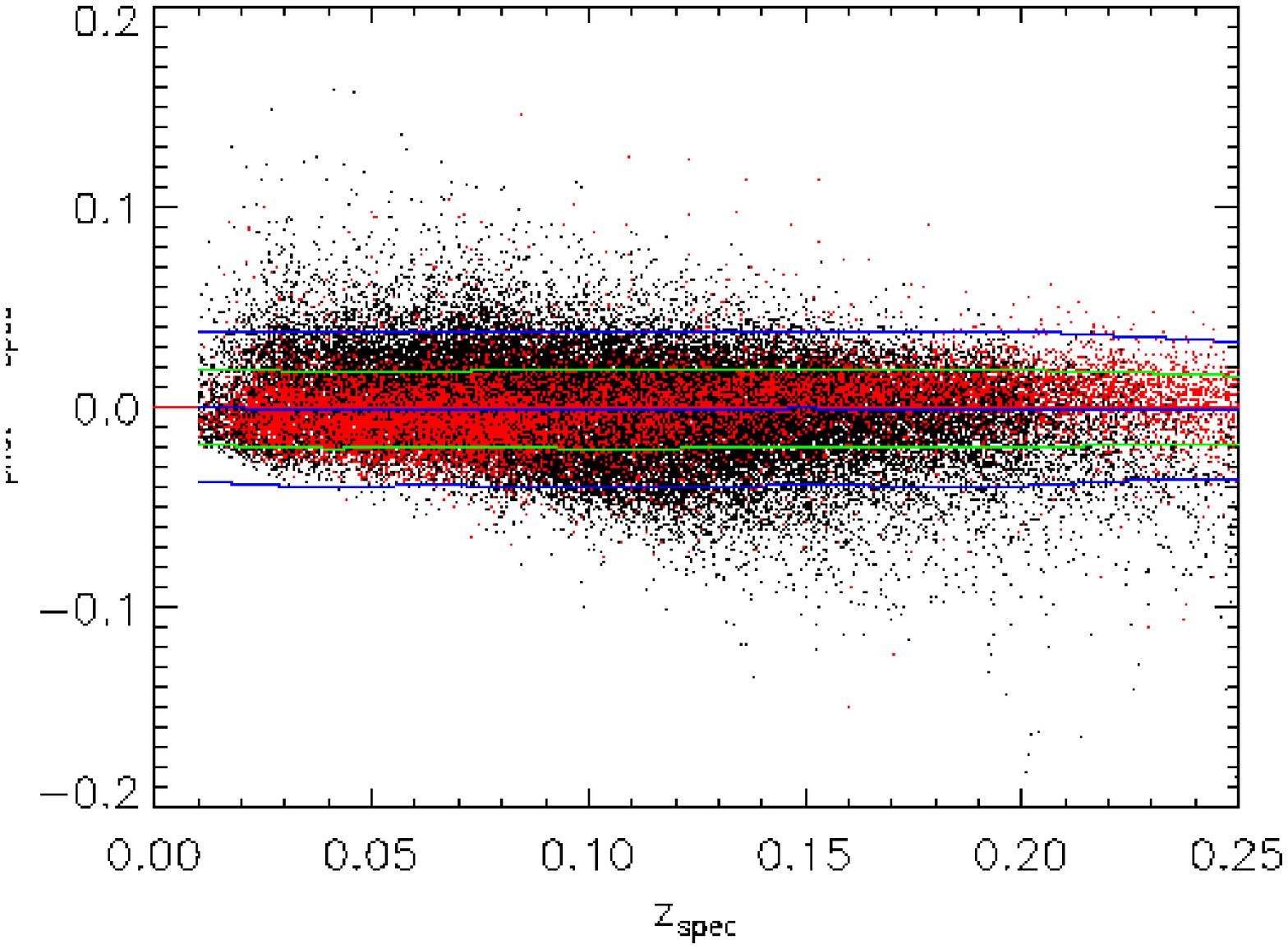}\includegraphics[angle=90,width=5cm,keepaspectratio]{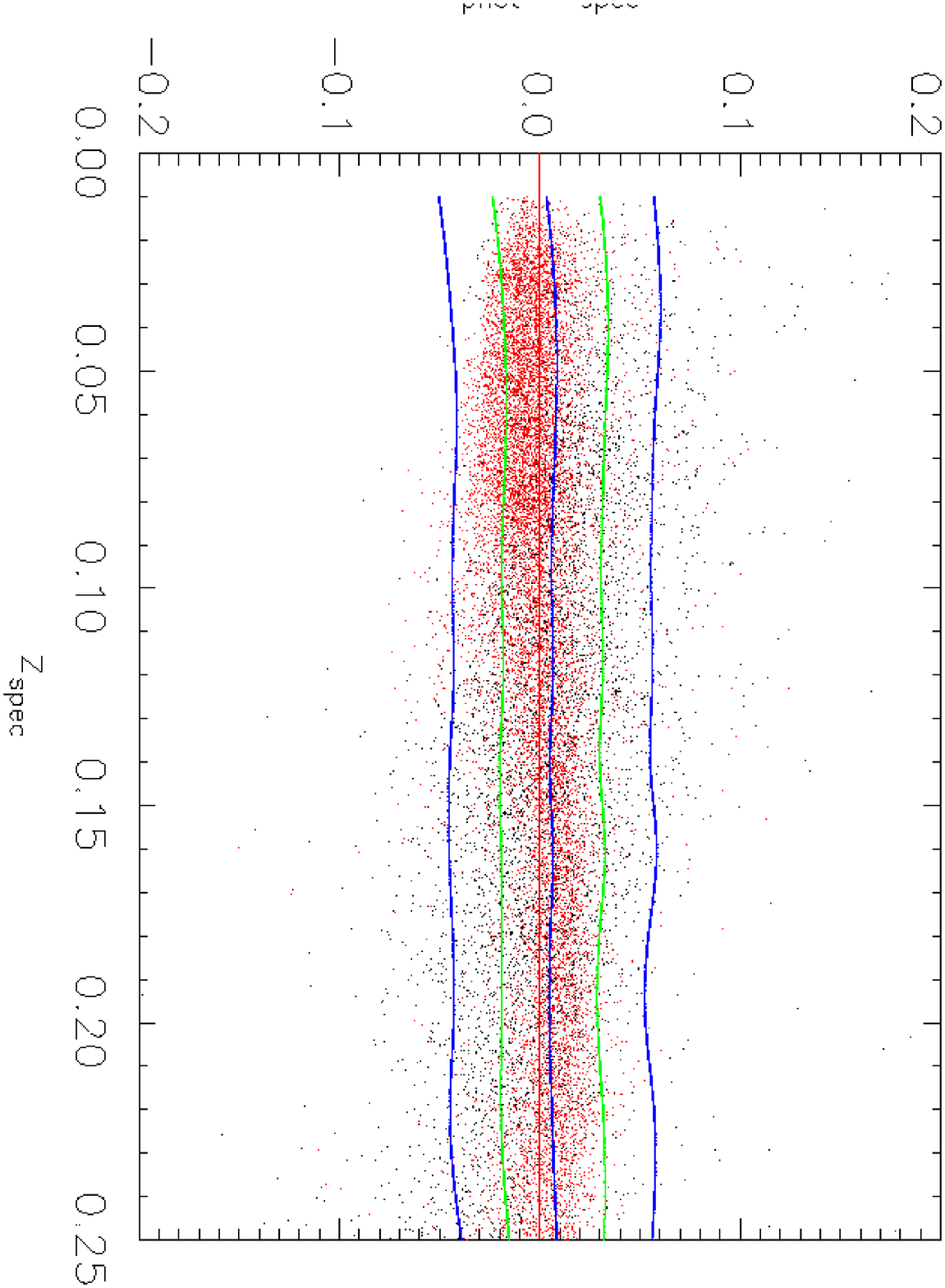}}
\centerline{\includegraphics[angle=90,width=5cm,keepaspectratio]{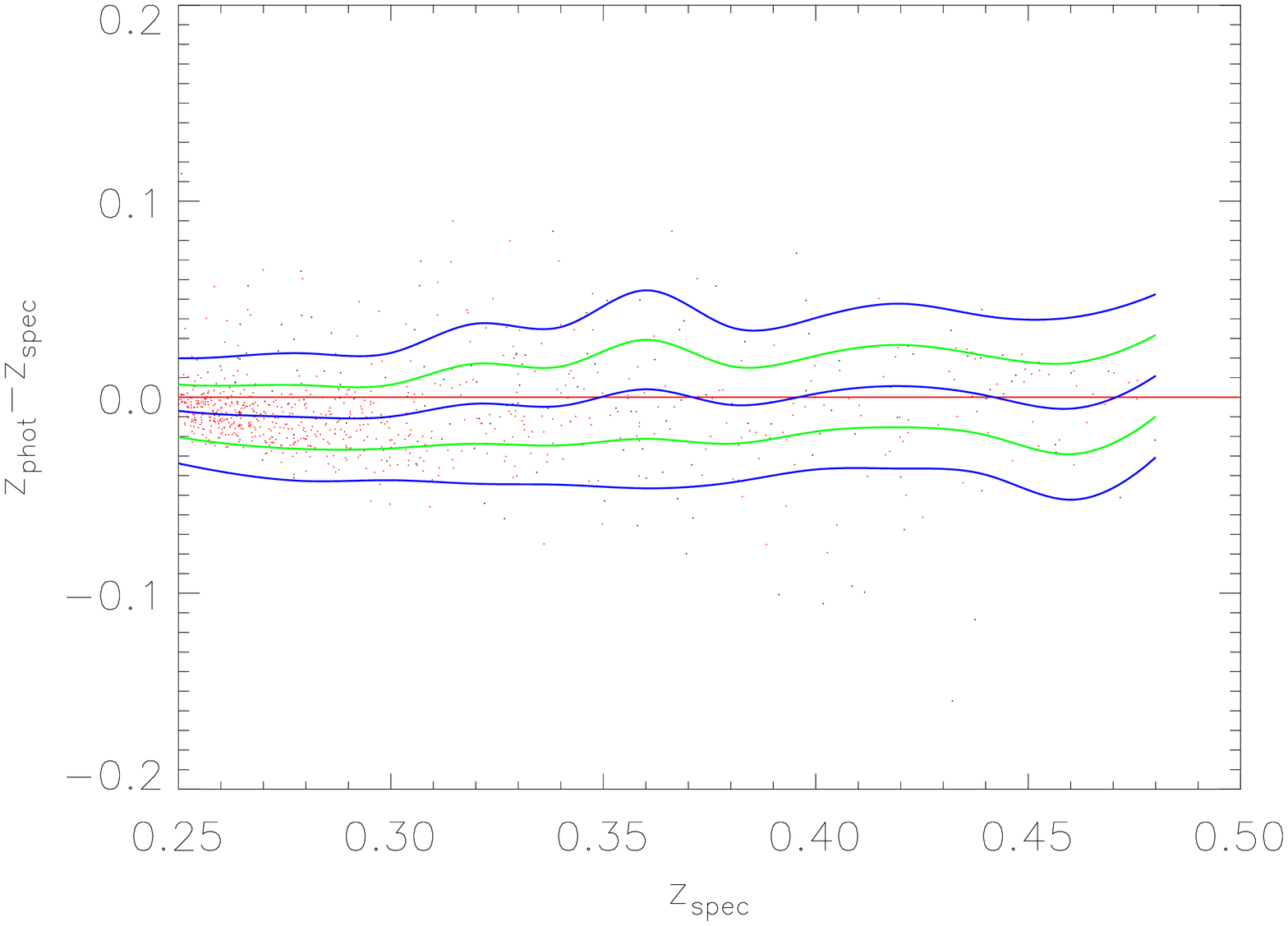}\includegraphics[angle=90,width=5cm,keepaspectratio]{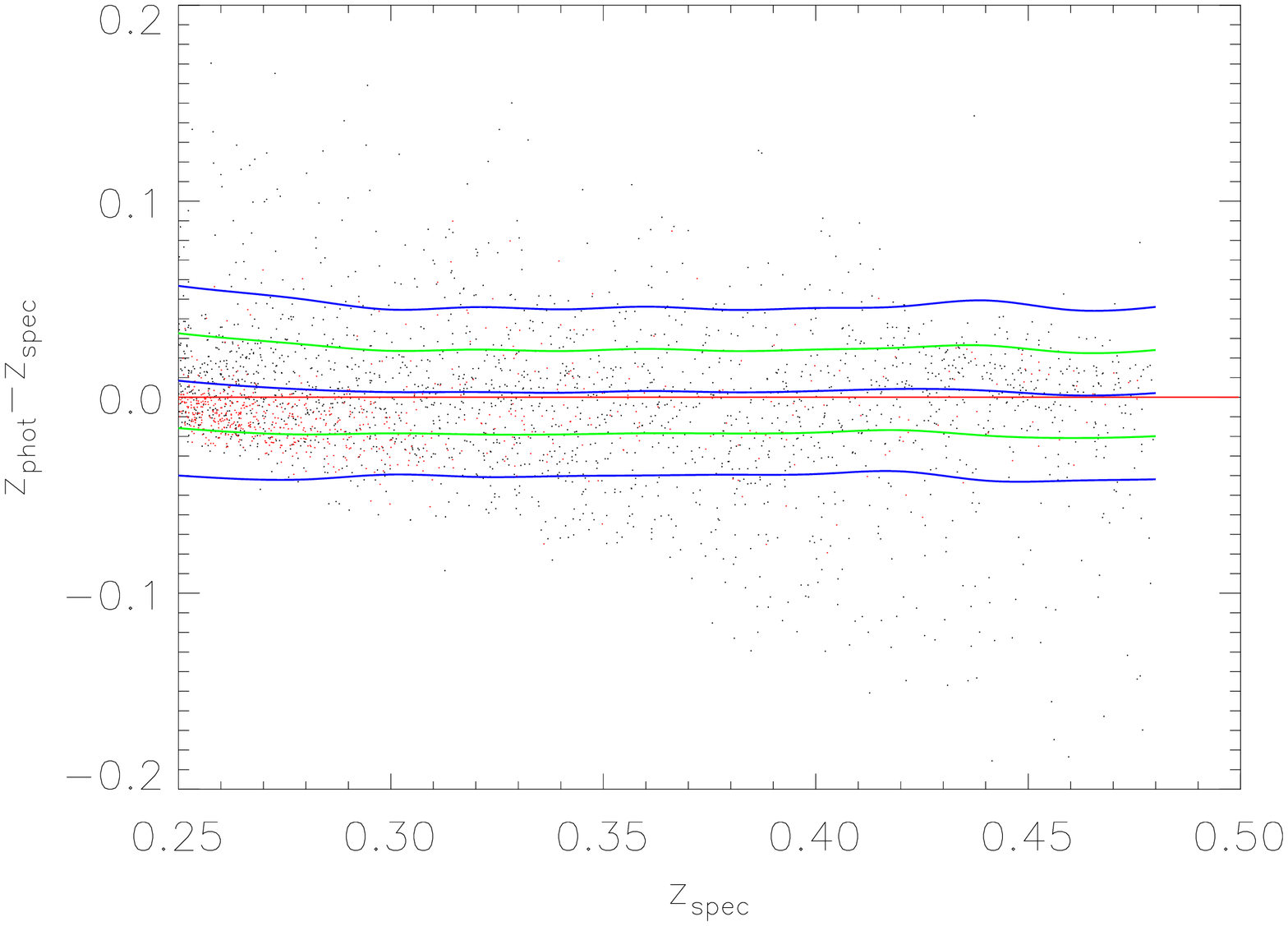}}
\caption{
Distribution of residuals for the GG sample divided in magnitude bins.
Upper left panel: nearby sample, $r<17.7$; upper right panel: nearby sample, $r>17.7$;
Lower left panel: distant sample, $r<17.7$; lower right panel: $17.7<r$.}
\label{Fig:GG_mag_bins}
\end{figure}

\begin{figure}
\centerline{\includegraphics[angle=90,width=5cm,keepaspectratio]{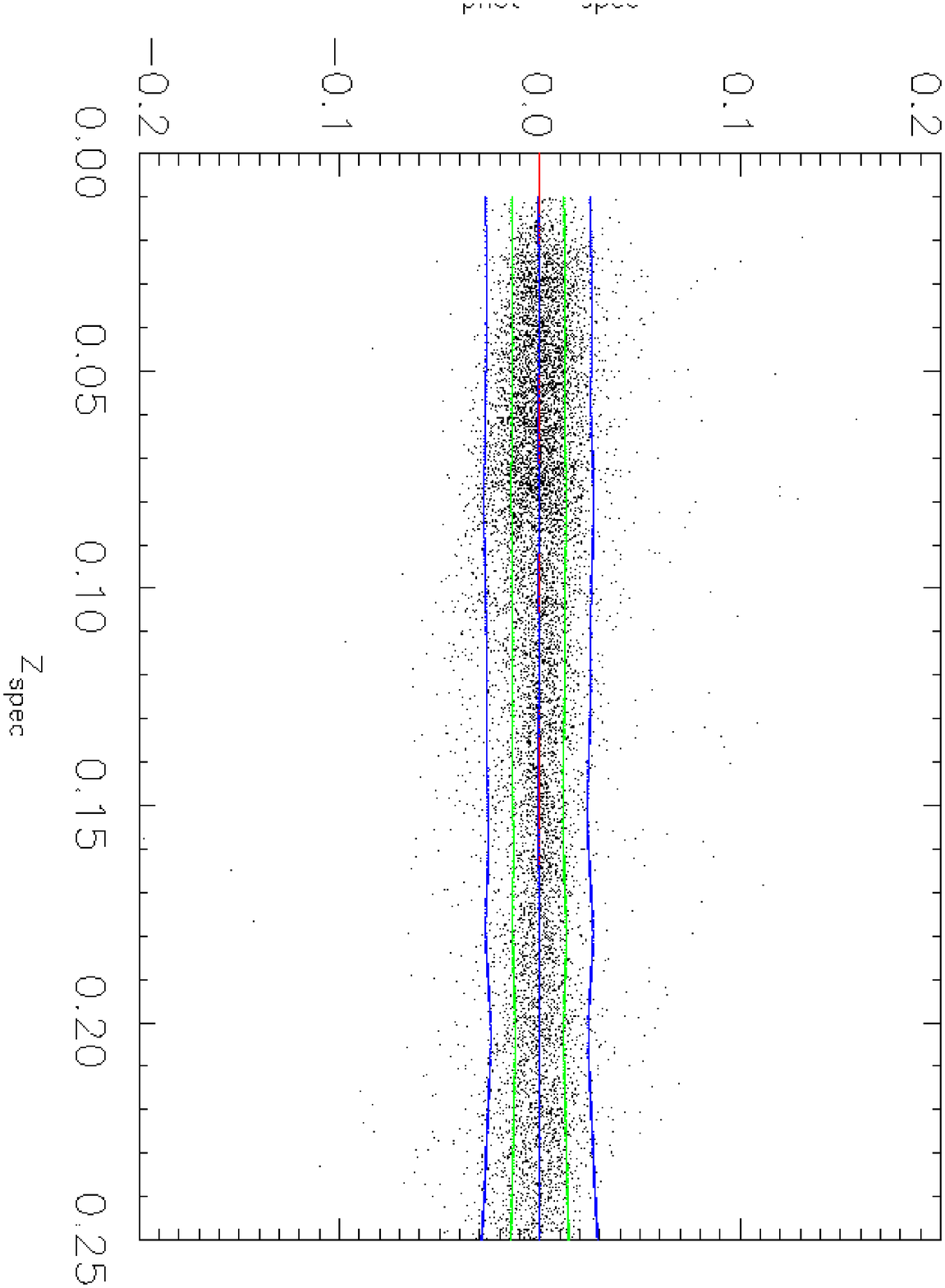}
\includegraphics[angle=90,width=5cm,keepaspectratio]{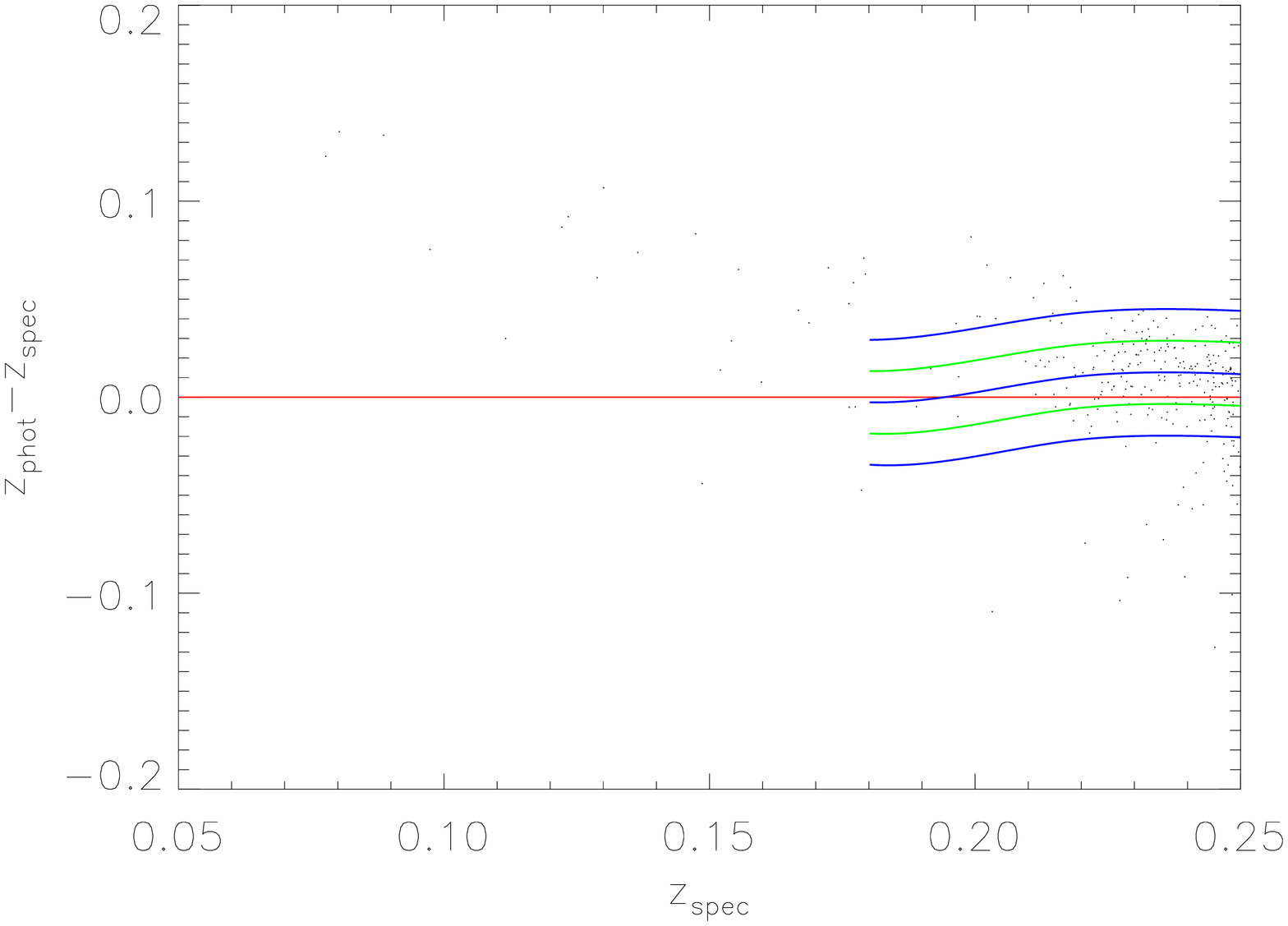}}
\centerline{\includegraphics[angle=90,width=5cm,keepaspectratio]{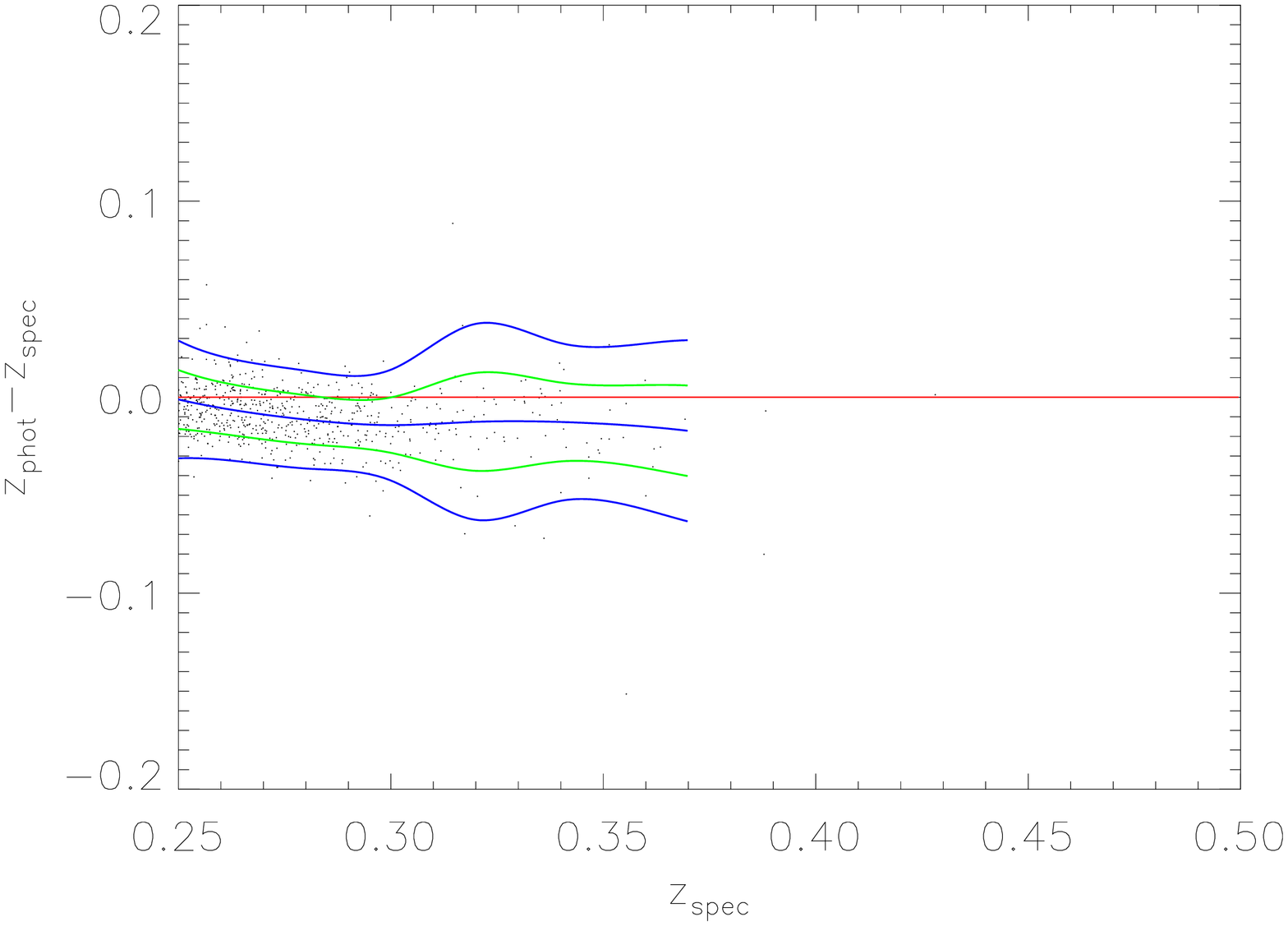}\includegraphics[angle=90,width=5cm,keepaspectratio]{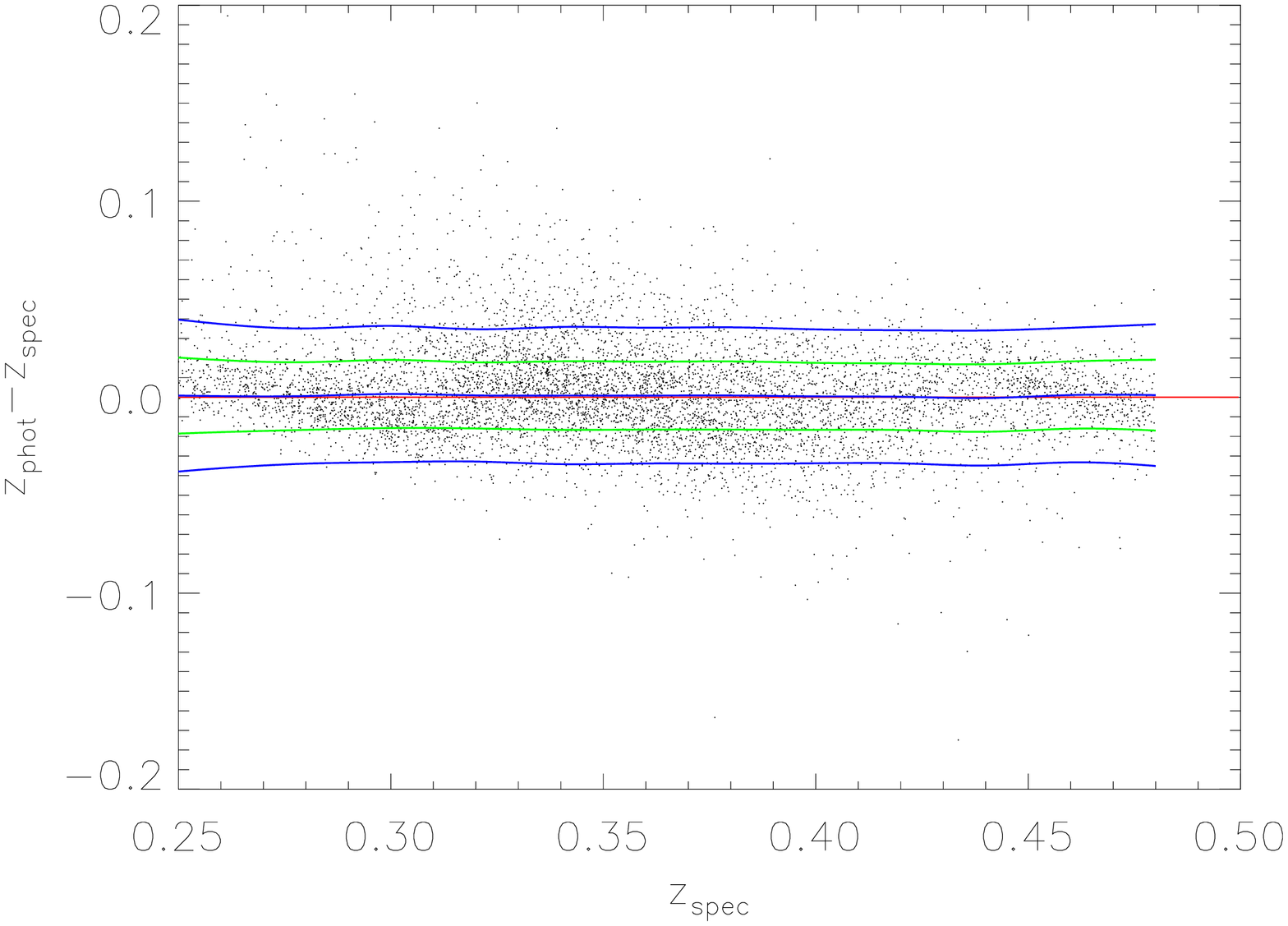}}
\caption{Distribution of residuals for the LRG sample divided in magnitude bins.
Upper left panel: nearby sample, $r<17.7$; upper right panel: nearby sample, $17.7<r$;
Lower left panel: distant sample, $r<17.7$; lower right panel: $17.7<r$.}
\label{Fig:LRG_mag_bins}
\end{figure}

The distribution of the residuals as a function of the spectroscopic redshift
for the GG and LRG samples is shown in Fig.~\ref{Fig:errors_totale} separately for the
near and distant objects.
We have also studied the dependance of such residuals from the
$r$-band luminosity of the galaxies in the two different magnitude ranges (cf. \ref{experiments}),
($r < 17.7$ and $r > 17.7$) and in the near and intermediate redshift bins, as shown in
Fig.~\ref{Fig:GG_mag_bins} and Fig.~\ref{Fig:LRG_mag_bins}
for the GG and LRG galaxies respectively.
Clear systematics are found only for near/faint and intermediate/luminous LRGs residuals:
in the former case, the mean value of residual $z_{phot} - z_{spec}$ is systematically higher then 0,
while in the latter it is costantly biased to negative values.
Both cases can be addressed reminding that these galaxies occupy a poorly sampled
volume in the parameter-space, and therefore the NN fails to reproduce the exact trend
of spectroscopic redshift.

\begin{figure}
\centering
   \resizebox{\hsize}{!}{\includegraphics[angle=90,keepaspectratio]{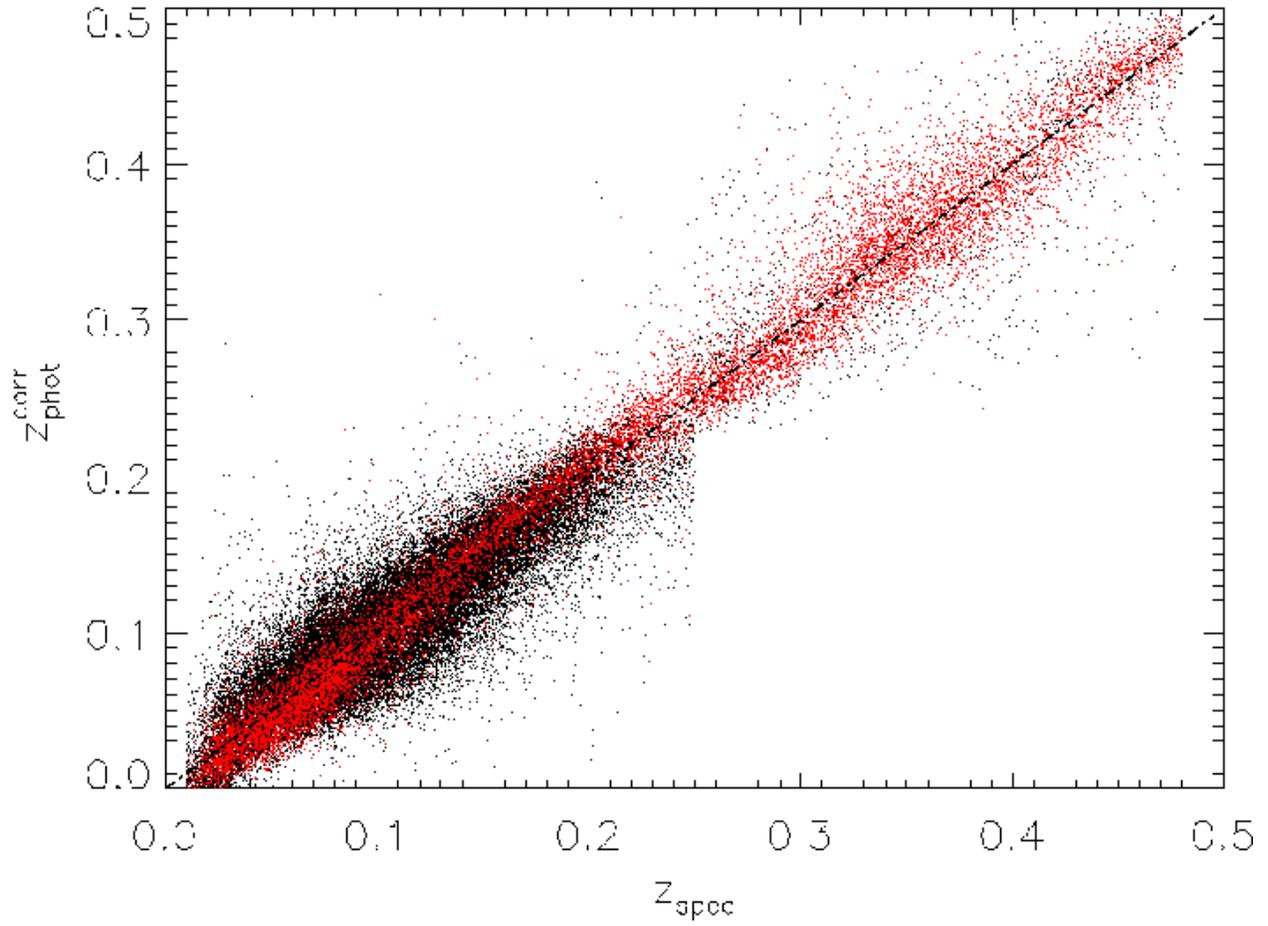}}
\caption{Plot of the same data shown in the lower panel of Fig.~\ref{Fig:GGS_omc}, with the
LRG objects marked as red dots. }
\label{Fig:LRG_GG}
\end{figure}

In Fig.~\ref{Fig:LRG_GG} we show the same plot as in Fig.~\ref{Fig:GGS_omc} but without isocontours and plotting as
red dots the objects which ''a posteriori'' were labeled as members of the LRG sample.
Interestingly enough, in the nearby sample the non-LRG and the LRG have robust variances
of $\sigma_3 = 0.021$ and $\sigma_3 =0.020$. Notice, however, that
the LRG objects show a clear residual systematic trend.
This behaviour can be explained by the fact that
in the nearby sample the training set contains a large enough
number of examples for both samples of objects and the network
can therefore achieve a good generalization capability.
In the distant sample the Non-LRG and LRG objects
have instead robust variances given by: $\sigma_3 = 0.321$ and $\sigma_3=0.021$.
Also in this case the observed behavior can be easily explained as
due to the heavy bias toward the LRGs which form $\sim 88.5 \%$ of the sample.
It must be stressed that while the remaining $11.4\%$ of the objects still
constitute a fairly large sample of objects, the uneven distribution of the training data
between the two groups of objects, overtrains the NN toward the LRG
objects which therefore are much better traced.

This confirms what already found by several authors
\citep{padmahaban_2005}: the derivation of photometric redshifts
requires besides than an accurate evaluation of the errors also
the identification of an homogeneous sample of objects.
\clearpage
\clearpage
\label{catastrophic}
Objects not matching the $3\sigma$ criterium used for the robust variance are:
3.47$\%$ for the LRG sample and $3.18\%$ for the GG sample.
Before correction, the rejected points are $\simeq 2\%$ of the
overall distribution for the GG sample and $\simeq 1.8\%$ for the LRG one.

As it was already mentioned, the SDSS data set has been
extensively analyzed by several authors who have used different
methods for photometric redshift determination. Unfortunately, a
direct comparison is not always possible due to differences in
either the data sets (different data releases have been used) or
in the way errors were estimated. It must be stressed, however,
that due to the fact that, above a minimum and reasonably low
treshold, the NN performances are not affected much by the number
of objects in the training set, the former factor can be safely
neglected. So far, the most extensive works are those by
\citep{csabai_2003} and \cite{way_2006}. In the former various
methods were tested against the EDR data. With reference to their
Table 3, and using the 'iterated' $\sigma$ which almost coincides
with the robust variance adopted here, we find that the best
performances were obtained, among the SED fitting methods for the
BC synthetic spectra ($\sigma_{it} \simeq 0.0621$ and $\sigma_{it}
\simeq 0.0306$, for the GG and LRG samples respectively. This
method, however leads to very clear systematic trends and to a
large number of catastrophic outliers ($\sim 3.5\%$). Much better
performances were attained by empirical methods and, in
particular, by the interpolative one which leads to a
$\sigma_{it}\simeq 0.0273$) with a fraction of catastrophic
redshifts of only $2\%$. In \cite{way_2006} the authors made use
of an Ensemble of NN (E) and Gaussian Process Regressions (GP).
Their best results using the magnitudes only were $0.0205$ and
$0.0230$ for the E and GP methods respectively, and at a
difference with our method, their methods greatly benefits by the
use of additional parameters such as the Petrosian radii, the
concentration index and the shape parameter. 

Two points are worth to be stressed. First of all, their selection criteria for the
construction of the training set appear much more restrictive and
it is not clear what performances could be achieved should such
restriction be relaxed. Second, even though such 'ensemble'
approach is very promising and is likely to be the most general
one, it has to be stressed that the bagging procedure, used in
\cite{way_2006} to combine the NNs, is known to be very effective
only in those cases where the intrinsic variance of the adopted
machine learning model is high. In this specific case, large
number of training data and few input features, the NN result very
stable and therefore other combining procedures, such as AdaBoost
(\cite{freund_1996}), should be preferred
(\cite{dietterich_2002}). This might also be the reason why when
only the photometric parameters are used their method gives
slightly worse performances than ours and instead leads to better
results when the number of features is increased.
\clearpage

\clearpage

An additional machine learning approach, namely Support Vector Machines, was
used by \cite{wadadekar_2004}. In Table \ref{comparison} we shortly summarize the
main results of the above quoted papers.

\begin{figure}
\centerline{\includegraphics[width=10cm,keepaspectratio]{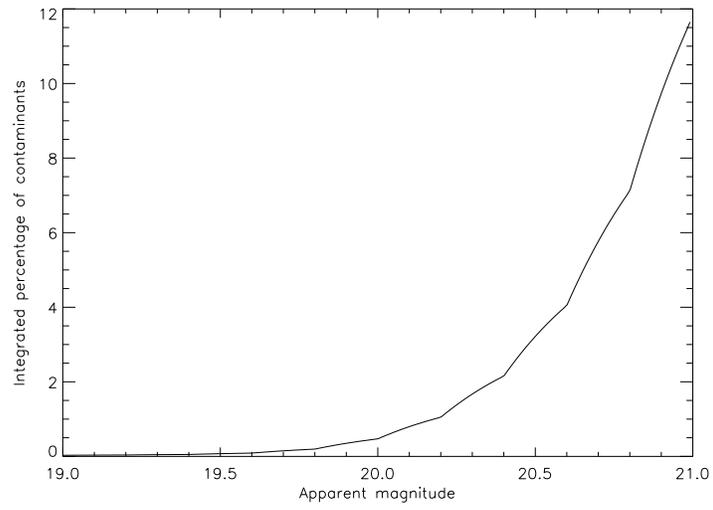}}
\caption{Estimated distribution of contaminants as a function of the apparent $r$ magnitude.
The $y$-axis gives the expected fraction of objects at $z>0.5$ which are erroneously evaluated by our procedure. }
\label{Fig:contaminants}
\end{figure}

\subsection{Contamination by distant galaxies}

The fact that our NN's are trained on a sample of galaxies with observed redshift 
$z_{spec} < 0.5$ introduces some contamination from objects which even though 
at $z>0.5$ still have $r<21$ and therefore match the photometric selection criteria. 

The only possible way to avoid such an effect would be to use a knowledge base covering 
in an uniform way all significant regions of the photometric parameter space down to the 
adopted magnitude limit.
In the case of SDSS this is true for magnitudes brighter than 17.7 but is not true at 
fainter light levels where the only region uniformously covered by the spectroscopic 
subsample is that defined by the LRG selection criteria.
A possible way out could be to extend the base of knowledge to fainter light level by 
including stastically significant and complete samples of spectroscopic redshifts from 
other and deeper surveys.
The feasibility of using a third NN to classify (and eventually throw into a waste basket) 
objects having $z>0.5$ is under study. At the moment, however, since we are interested in 
validating the method and in producing catalogues to be used for statistical applications, 
we shall estimate the number and the distribution in magnitude of such contaminants on 
statistical grounds only using the $r$-band luminosity function derived from SDSS data 
by \cite{blanton_2003}. This function, in fact, allows to derive for any given absolute 
magnitude the number of objects which even though at a redshift larger than 0.5
still match our apparent magnitude threshold and thus are misclassified.
By integrating over the absolute magnitude and over the volume
covered by the survey we obtain the curve in Fig. \ref{Fig:contaminants} 
which corresponds to a total number of
contaminants of $\sim 3.74 \times 10^6$. It has to be noticed however that 
for magnitudes brighter than 20.5, the fraction of contaminants is less than 
$0.04$ and drops below $0.01$ for $r<20$.

\clearpage
%\begin{deluxetable*}{llllll}
\begin{deluxetable}{llllll}
\tabletypesize{\scriptsize}
\def\arraystretch{1.0}
\tablecolumns{6}
\tablewidth{0pt}
\tablecaption{Comparisons of various methods for the photometric redshift estimation
applied to the SDSS data.}
\tablehead{
\colhead{Reference}  &\colhead{Method}  &\colhead{Data}  &\colhead{$\Delta z$}  &\colhead{$\sigma$}  &\colhead{Range}
}
\startdata
\cite{csabai_2003}    & SED fitting CWW             & EDR     &       &0.0621    &           \\
\cite{csabai_2003}    & SED fitting BC              & EDR     &       &0.0509    &           \\
\cite{csabai_2003}    & interpolative               & EDR     &       &0.0451    &           \\
\cite{csabai_2003}    & bayesian                    & EDR     &       &0.0402    &           \\
\cite{csabai_2003}    & empirical, polynomial fit   & EDR     &       &0.0318    &           \\
\cite{csabai_2003}    & K-D tree                    & EDR     &       &0.0254    &           \\
\cite{classx}         & Class X                     & DR-2    &       &0.0340    &           \\
\cite{way_2006}$^{\mathrm{a}}$       & Gaussian Process            & DR-3    &       &0.0230    &           \\
\cite{way_2006}$^{\mathrm{a}}$       & ensemble                    & DR-3    &       &0.0205    &           \\
\cite{collister_2004} & ANNz                        & EDR     &       &0.0229    &           \\
\cite{wadadekar_2004} & SVM                         & DR-2    &       &0.027     &           \\
\cite{wadadekar_2004}$^{\mathrm{a}}$ & SVM                         & DR-2    &       &0.024     &\\
\cite{vanzella_2003}  & MLP ff                      & DR1     &  0.016&0.022     & $<0.4$\\
\cite{padmahaban_2005}& Template fitting and Hybrid & DR1-LRG & $<0.01$&$\sim 0.035$& $<0.55$\\
this work before int. & MLP                         & DR5-GG  &-0.0036&0.0197    & 0.01,0.25 \\
this work before int. & MLP                         & DR5-GG  &-0.0036&0.0245    & 0.25,0.48 \\
this work after  int. & MLP                         & DR5-GG  &       &          & 0.01,0.25 \\
this work after  int. & MLP                         & DR5-GG  &       &          & 0.25,0.48 \\
this work before int. & MLP                         & DR5-LRG &-0.0029&0.0194    & 0.01,0.25 \\
this work before int. & MLP                         & DR5-LRG &-0.0029&0.0205    & 0.25,0.48 \\
\enddata
\tablecomments{
Column 1: reference;
Column 2: method (for the acronyms see text);
Column 3: data set (EDR=Early Data Release; DR1 through DR5 the various SDSS data release);
Column 4: systematic offset;
Column 5: standard deviation;
Column 6: redshift range over which the average error is estimated.\\
($^{\mathrm{a}}$): additional morphological and photometric
parameters. }

\label{comparison}
%\end{deluxetable*}
\end{deluxetable}
\clearpage
\section{The catalogues}
\label{catalogues} As mentioned above, the catalogues containing the
photometric redshift parameters together with the parameters used
for their derivation can be downloaded at the URL:
$\mathrm{http://people.na.infn.it/~astroneural/SDSSredshifts.htm/}$.
This data, for consistency with the SDSS survey, has been subdivided in several files,
each corresponding to a different SDSS \emph{stripe} of the observed
sky. A \emph{stripe} is defined by a line of constant
survey latitude $\eta$, bounded on the north and south by the edges
of the two strips (scans along a constant $\eta$ value), and bounded
on the east and west by lines of constant lambda. Because both
strips and stripes are defined in "observed" space, they are
rectangular areas which overlap as one approaches the poles (for
more details see $\mathrm{http://www.sdss.org}$). The data for both GG and 
LRG samples have been extracted using the queries described in ~\ref{thedata}.
\noindent The catalogues can be downloaded as 'FITS' files,
containing the fundamental parameters used for
redshift determination and the estimated
photometric redshift for each individual source. In more details (in brackets SDSS database names of the parameters):
unique SDSS identifier ('objID'),right ascension J2000 ('ra')
declination J2000 ('dec'), dereddened magnitudes ('dered\_u', 'dered\_g', 'dered\_r', 'dered\_i', 'dered\_z'), radius containing 50$\%$ of Petrosian flux for each magnitude ('petroR50\_u', 'petroR50\_g', 'petroR50\_r', 'petroR50\_i', 'petroR50\_z'), radius containing 90$\%$ of Petrosian flux for each magnitude ('petroR90\_u', 'petroR90\_g', 'petroR90\_r', 'petroR90\_i', 'petroR90\_z'),
De Vaucouleurs fit ln(likelihood) in u and r bands ('lnLDeV\_u', 'lnLDeV\_r'),
exponential disc fit ln(likelihood) in u and r bands ('lnLExp\_u', 'lnLExp\_r').

\section{Conclusions}\label{conclusion}

In the previous paragraphs we discussed a 'two steps'
application of neural networks to the evaluation of photometric redshifts.
Even though finely tailored on the characteristic of the SDSS, the method
is completely general and can be easily applied to any other multiband
data set provided that a suitable base of spectroscopic knowledge
is available.
As most other neural networks methods, several advantages
are evident:

\begin{enumerate}
\item The NN can be easily re-trained if new data become available.
Even thoug the training phase can be rather demanding in terms of
computing time, once the NN has been trained, the derivation of redshifts
is almost immediate ($10^7$ objects are processed on the fly on a normal
laptop).
\item Even though it was not necessary in this specific case, all sorts of
a priori knowledge can be taken into account.
\end{enumerate}
On the other end, the method suffers of those limitations which are typical of all
empirical methods based on interpolation.
Most of all, the training set needs to 
ensure a complete and if possible uniform coverage of the
parameter space.

Our method allowed to derive photometric redshifts for $z\lesssim 0.5$ with robust variances
of $\sigma_3 =0.0208$ for the GG sample ($\sigma_3 =0.0197$ and $\sigma_3 =0.0238$ for the nearby and distant sample respectively) and $\sigma_3 =0.0164$ for the LRG sample ($\sigma_3 =0.0160$ and $\sigma_3 =0.0183$). This accuracy was reached adopting using a two-step approach
allowing to build training sets which uniformly sample the parameter space of the overall population.

In the case of LRGs, the better accuracy and the close Gaussianity of the residuals, are explained by the fact that this sample was selected based on the a priori assumption that they form a rather homogeneous population sharing the same SED. 
In other words, this result confirms what has long been known, id est the fact that
when using empirical methods, it is crucial to define photometrically homogeneous
populations of objects.

In the more general case it would be necessary to define photometrically homogeneous
populations of objects in absence of a priori information and therefore relying only on the
photometric data themselves.
This task, as it has been shown for instance by \cite{classx,bazell_2004} is a non trivial one,
since the complexity of astronomical data and the level of degeneration is so high that most
unsupervised clustering methods partition the photometric parameter space in far too many clusters, thus preventing the build-up a of a suitable base of knowledge.
A possible way to solve this problem will be discussed in Paper III.

\acknowledgements
This work was supported by the European VO-Tech {European Virtual Observatory
Technological Infrastructure} project  and by the MIUR-PRIN program.
The authors also thank the Regione Campania for partial financial Support.
We thank Gennaro Miele and Giampiero Mangano for useful discussions.

\label{lastpage}
\end{document}